\documentclass{aa}  

\usepackage{graphicx}
\usepackage{txfonts}

\newcommand\ztwo{2\,$<$\,$z$\,$<$\,2.5}
\newcommand\lya{Ly$\alpha$}
\newcommand{\etal}{{et\thinspace al.}}
\newcommand{\Ho}{$H_{\rm 0}$}
\newcommand{\tabref}[1]{Table~\ref{#1}}
\newcommand{\figref}[1]{Figure~\ref{#1}}
\newcommand{\secref}[1]{\S\ref{#1}}
\newcommand\sfgn{SFG$_{\rm N}$}
\newcommand\sfgl{SFG$_{\rm L}$}
\newcommand\rsc{$\rho$}

\begin{document} 

\title{The VIMOS Ultra Deep Survey: \lya\ Emission and Stellar
  Populations of Star-Forming Galaxies at \ztwo\ \thanks{Based on data
    obtained with the European Southern Observatory Very Large
    Telescope, Paranal, Chile, under Large Program 185.A-0791.}}

\author{N. P. Hathi\inst{1}
\and O. Le F\`evre\inst{1}
\and O. Ilbert\inst{1}
\and P. Cassata\inst{18}
\and L. A. M. Tasca\inst{1}
\and B. C. Lemaux \inst{1}
%
\and B. Garilli\inst{3}
\and V. Le Brun\inst{1}
\and D. Maccagni\inst{3}
\and L. Pentericci\inst{4}
\and R. Thomas\inst{1}
\and E. Vanzella\inst{2}
\and G. Zamorani \inst{2}
\and E. Zucca\inst{2}
%
\and R. Amor\'in\inst{4}
\and S. Bardelli\inst{2}
\and L.P. Cassar\`a\inst{3}
\and M. Castellano\inst{4}
\and A. Cimatti\inst{5}
\and O. Cucciati\inst{5,2}
\and A. Durkalec\inst{1}
\and A. Fontana\inst{4}
\and M. Giavalisco\inst{13}
\and A. Grazian\inst{4}
\and L. Guaita\inst{4}
\and A. Koekemoer\inst{17}
\and S. Paltani\inst{9}
\and J. Pforr\inst{1}
\and B. Ribeiro\inst{1}
\and D. Schaerer\inst{10,8}
\and M. Scodeggio\inst{3}
\and V. Sommariva\inst{5,4}
\and M. Talia\inst{5}
\and L. Tresse\inst{1}
\and D. Vergani\inst{6,2}
%
\and P. Capak\inst{12}
\and S. Charlot\inst{7}
\and T. Contini\inst{8}
\and J.G. Cuby\inst{1}
\and S. de la Torre\inst{1}
\and J. Dunlop\inst{16}
\and S. Fotopoulou\inst{9}
\and C. L\'opez-Sanjuan\inst{11}
\and Y. Mellier\inst{7}
\and M. Salvato\inst{14}
\and N. Scoville\inst{12}
\and Y. Taniguchi\inst{15}
\and P.W. Wang\inst{1}
}
\institute{Aix Marseille Universit\'e, CNRS, LAM (Laboratoire d'Astrophysique de Marseille) UMR 7326, 13388, Marseille, France
\and
INAF--Osservatorio Astronomico di Bologna, via Ranzani,1, I-40127, Bologna, Italy
\and
INAF--IASF, via Bassini 15, I-20133, Milano, Italy
\and
INAF--Osservatorio Astronomico di Roma, via di Frascati 33, I-00040, Monte Porzio Catone, Italy
\and
University of Bologna, Department of Physics and Astronomy (DIFA), V.le Berti Pichat, 6/2 - 40127, Bologna, Italy
\and
INAF--IASF Bologna, via Gobetti 101, I--40129, Bologna, Italy
\and
Institut d'Astrophysique de Paris, UMR7095 CNRS, Universit\'e Pierre et Marie Curie, 98 bis Boulevard Arago, 75014 Paris, France
\and
Institut de Recherche en Astrophysique et Plan\'etologie - IRAP, CNRS, Universit\'e de Toulouse, UPS-OMP, 14, avenue E. Belin, F31400
Toulouse, France
\and
Department of Astronomy, University of Geneva, ch. d'\'Ecogia 16, CH-1290 Versoix, Switzerland
\and
Geneva Observatory, University of Geneva, ch. des Maillettes 51, CH-1290 Versoix, Switzerland
\and
Centro de Estudios de F\'isica del Cosmos de Arag\'on, Teruel, Spain
\and
Department of Astronomy, California Institute of Technology, 1200 E. California Blvd., MC 249--17, Pasadena, CA 91125, USA
\and
Astronomy Department, University of Massachusetts, Amherst, MA 01003, USA
\and
Max-Planck-Institut f\"ur Extraterrestrische Physik, Postfach 1312, D-85741, Garching bei M\"unchen, Germany
\and
Research Center for Space and Cosmic Evolution, Ehime University, Bunkyo-cho 2-5, Matsuyama 790-8577, Japan
\and
SUPA, Institute for Astronomy, University of Edinburgh, Royal Observatory, Edinburgh, EH9 3HJ, United Kingdom
\and
Space Telescope Science Institute, 3700 San Martin Drive, Baltimore, MD 21218, USA 
\and
Instituto de Fisica y Astronomia, Facultad de Ciencias, Universidad de Valparaiso, Av. Gran Bretana 1111, Casilla 5030, Valparaiso, Chile
}
   \date{Received  ??, 2014; accepted ??, 2014}

   \abstract{The aim of this paper is to investigate spectral and
     photometric properties of 854 faint
     ($i_{AB}$\,$\lesssim$\,25~mag) star-forming galaxies (SFGs) at
     \ztwo\ using the VIMOS Ultra-Deep Survey (VUDS) spectroscopic
     data and deep multi-wavelength photometric data in three
     extensively studied extragalactic fields (ECDFS, VVDS,
     COSMOS). These SFGs were targeted for spectroscopy based on their
     photometric redshifts. The VUDS spectra are used to measure the
     UV spectral slopes ($\beta$) as well as \lya\ equivalent widths
     (EW). On average, the spectroscopically measured $\beta$
       (--1.36$\pm$0.02), is comparable to the photometrically
       measured $\beta$ (--1.32$\pm$0.02), and has smaller
       measurement uncertainties. The positive correlation of $\beta$
     with the Spectral Energy Distribution (SED)-based
     measurement of dust extinction E$_{\rm s}$(B--V) emphasizes the
     importance of $\beta$ as an alternative dust indicator at high
     redshifts.  To make a proper comparison, we divide these SFGs into
     three subgroups based on their rest-frame \lya\ EW: SFGs
       with no \lya\ emission (\sfgn; EW\,$\le$\,0\AA), SFGs with
       \lya\ emission (\sfgl; EW\,$>$\,0\AA), and \lya\ emitters
       (LAEs; EW\,$\ge$\,20\AA).  The fraction of LAEs at these
     redshifts is $\sim$10\%, which is consistent with previous
     observations. We compared best-fit SED-estimated stellar
     parameters of the \sfgn, \sfgl\ and LAE samples.  For the
       luminosities probed here ($\sim$L$^*$), we find that galaxies
       with and without Ly$\alpha$ in emission have small but
       significant differences in their SED-based properties. We find
       that LAEs have less dust, and lower star-formation rates (SFR)
       compared to non-LAEs. We also find that LAEs are less massive
       compared to non-LAEs, though the difference is smaller and less
       significant compared to the SFR and E$_{\rm s}$(B--V).  When we
     divide the LAEs based on their Spitzer/IRAC 3.6$\mu$m fluxes, we
     find that the fraction of IRAC-detected
     (m$_{\rm 3.6}$\,$\lesssim$\,25~mag) LAEs is much higher than the
     fraction of IRAC-detected narrow band (NB)-selected LAEs at
     $z$\,$\simeq$\,2--3. This could imply that UV-selected LAEs host
     a more evolved stellar population, which represents a later stage of
     galaxy evolution compared to NB-selected LAEs. }


\titlerunning{VUDS: Ly$\alpha$ emission and Stellar Populations at  \ztwo}
\authorrunning{N.~P.\ Hathi et al.}

   \maketitle

\section{Introduction}

In the last decade, large numbers of star-forming galaxies (SFGs) ---
selected using the Lyman break color technique and/or the photometric
redshift technique --- were identified and studied at
$z$\,$\gtrsim$\,3 using deep photometric surveys
\citep[e.g.,][]{bouw07,hath08a,
  ouch09,mclu10,yan10,fink10,hath12,tilv13,elli13,fink13}. This large
reservoir of SFGs spanning a very large redshift range
($z$\,$\simeq$\,3--8), and brightnesses (24--29~mag), has tremendous
impact on our understanding of the process of galaxy formation and
evolution \citep[e.g.,][]{maio08,ilbe13,spea14,fink15}. However,
the spectroscopic and photometric studies for most of these galaxies
are very challenging because of their faint magnitudes and lack of
high resolution multi-wavelength data. Therefore, detailed studies of
SFGs at lower redshifts ($z$\,$<$\,3) using ground-based spectroscopy
and multi-wavelength photometry are vital for understanding physical
processes at the peak epoch of star formation as well as to better
understand high redshift galaxy formation.

The study of SFGs at $z$\,$\sim$\,2 has a significant impact on our
understanding of galaxy properties.  First, these SFGs are at
redshifts corresponding to the peak epoch ($z$\,$\sim$\,1--3) of the
global star formation rate (SFR) density
\citep[e.g.,][]{mada98,cucc12,mada14}, where $>$50\% of the stars in
the present-day Universe formed, and is a perfect epoch to study star
formation processes.  Second, this cosmologically interesting redshift
is also where we have access to a wealth of multi-wavelength data,
and rest-frame optical spectroscopy, to study galaxy properties
that we cannot investigate at higher redshifts with current
technology. The major advantage of identifying and studying various
physical properties --- including star formation properties --- of
these SFGs is that they can be investigated in rest-frame UV as well
as rest-frame optical filters.  Third, these galaxies are likely lower
redshift analogs of the high redshift SFGs, and putting in a context
such analysis will help to shed light on the process of reionization
in the early Universe \citep[e.g.,][]{labb10,star10} and how the first
galaxies formed. Finally, the redshift $z$\,$\sim$\,2 is the lowest
redshift where we can identify and study \lya\ in emission from
ground-based spectroscopy and these galaxies are bright enough for
extensive spectroscopic studies on 8-10m class telescopes.

Spectroscopy is an essential and powerful tool to fully understand
galaxy properties but the underlying photometric selection of galaxies
can bias the spectroscopic sample properties.  The primary techniques
to select SFGs by color at $z$\,$\simeq$\,2 are: (1) \emph{sBzK}
\citep[using the $B$, $z$, $K$ bands,][]{dadd04,dadd07}, (2) BX/BM
\citep[using the $U$, $G$, $R$ bands,][]{stei04, adel04}, and (3) LBG
\citep[using the bands which bracket the redshifted Lyman
limit,][]{hath10,oesc10}. The other main approach is based on the
magnitude/flux limit and/or photometric redshift selection, as used by
many spectroscopic surveys such as, VVDS, GMASS, zCOSMOS
\citep[e.g.,][]{lefe05,lill07, kurk09,lefe13}.  On the other hand,
SFGs are also selected based on the emission-line/Narrow-band (NB)
techniques \citep[e.g.,][]{guai10,berr12,varg14} which bracket strong
emission lines at $z$\,$\simeq$\,2. All these approaches select
star-forming galaxies, and yield insight into the star-forming
properties of these galaxies, but they have differing selection
biases, and therefore, these samples do not completely overlap
\citep[see][for details]{ly11,habe12}. Therefore, it is essential to
apply a well defined selection leading to the identification of a
population with a broad range of galaxy properties, which can then be
compared to similar galaxies at higher redshifts.

The comparison of stellar populations of (strong) Ly$\alpha$
emitting galaxies (LAEs), primarily selected based on the NB imaging
technique, and non-LAEs (with weak or no \lya-emission), primarily
selected based on photometric colors that mimic a break (i.e., Lyman
break galaxies, LBGs), has yielded diverse and inconclusive
results. \citet{gawi06} found that NB-selected LAEs at
$z$\,$\simeq$\,3.1 are less massive and less dusty compared to
continuum selected LBGs suggesting that LAEs represent an early stage
of an evolutionary sequence where galaxies gradually become more
massive and dusty due to mergers and star formation
\citep{gawi07}. This conclusion is consistent with high specific star
formation rates (SSFR, SFR per unit mass) of NB-selected LAEs found by
\citet{lai08} relative to LBGs at the same redshift. On the other
hand, \citet{fink09} found a range of dust extinctions in a sample of
14 NB-selected LAEs at $z$\,$\simeq$\,4.5, which is not consistent
with LAEs being the first galaxies in the evolutionary
sequence. Similarly, \citet{nils11} used 171 NB-selected LAEs at
$z$\,$\simeq$\,2.25 and concluded that the stellar properties of LAEs
are different from those at higher redshift and that they are
diverse. They also believe that \lya\ selection could be tracing
different galaxies at different redshifts, which is consistent with
the findings of \citet{acqu12}. \citet{lai08} found that IRAC-detected
LAEs are significantly older (age\,$\gtrsim$\,1~Gyr) and more massive
(mass\,$\sim$\,10$^{10}$~M$_{\odot}$) compared to IRAC-undetected LAEs at
$z$\,$\simeq$\,3.1, which lead them to suggest that the IRAC-detected
LAEs may be a lower-mass extension of the LBG population. These
studies show heterogeneity of NB-selected LAEs, and comparison between
these objects and continuum selected LBGs continues to be the subject
of many new studies.

Various authors have compared stellar population properties of LAEs
and non-LAEs in UV flux-limited samples
\citep[e.g.,][]{shap01,erb06,pent07,redd08,korn10}. \citet{shap01}
generated rest-frame UV composite spectra of LBGs at $z$\,$\simeq$\,3
dividing subsamples in `young' ($t\le$\,35~Myr) and `old' ($t\ge$\,1~Gyr)
galaxies. They found that younger galaxies have weaker \lya\ emission,
are more dusty, and less massive compared to older galaxies.  On the
other hand, the analysis of \citet{erb06} of the composite UV spectra
for $\sim$60 SFGs at $z$\,$\simeq$\,2 concluded that, on average,
objects with lower stellar mass had stronger \lya\ emission line than
more massive objects.  Using a sample of 14 UV-selected galaxies at
$z$\,$\simeq$\,2--3 with \lya\ equivalent width (EW) $\ge$\,20\AA,
\citet{redd08} found no significant difference in the stellar
populations of strong \lya-emitters compared to the rest of the
sample.  At higher redshifts, \citet{pent07} found that, in general,
younger galaxies at $z$\,$\simeq$\,4 showed \lya\ in emission while
older galaxies showed \lya\ in absorption. More recently, \citet{korn10}
used 321 LBGs at $z$\,$\simeq$\,3 to study the relationship between
\lya-emission and stellar populations. Based on their analysis, they
concluded that objects with strong \lya-emission are older, lower in
SFR, and less dusty compared to objects with weak or no (\lya\ in
absorption) \lya-emission.  The results of these studies emphasize
that the exact relation between stellar populations and \lya-emission
is still not yet fully understood.

In this paper we use spectroscopic data from the VIMOS Ultra-Deep
Survey (VUDS), to study the spectral \& photometric properties of a large
sample of faint SFGs at \ztwo\ ($z_{median}$\,$\simeq$\,2.3). VUDS recently
obtained data for $\sim$10000 galaxies over an area of $\sim$1 deg$^2$
using the 8.2m Very Large Telescope (VLT).  The VUDS observations were
performed in the well-studied COSMOS \citep{scov07,koek07} and ECDFS
\citep{giav04,rix04} fields now partly covered with HST/WFC3 by
CANDELS \citep{grog11,koek11}, and in the VVDS-02h field
\citep{lefe04,mccr03}. These fields have extensive multi-wavelength
data.  We extract a sample of 854 galaxies with confirmed
spectroscopic redshifts at \ztwo, uniformly selected from their
continuum properties with $i_{AB}$\,$\lesssim$\,25~mag. The goal of this
paper is to study the physical properties of these SFGs which include
galaxies with and without \lya\ emission.

This paper is organized as follows: In \secref{data}, we summarize the
VUDS observations, and discuss \ztwo\ sample selection as well as
  its basic properties. In \secref{uvslope}, we fit observed
rest-frame UV VUDS spectra at \ztwo\ to measure the UV spectral slope,
and discuss its correlation with the E$_{\rm s}$(B--V), stellar mass,
and UV absolute magnitude.  In \secref{lae}, we discuss differences in
stellar population properties between galaxies with and without \lya\
in emission. In \secref{lae_sp}, we investigate correlations between
derived physical parameters (UV absolute magnitude, SFRs,
E$_{\rm s}$(B--V), stellar mass, and UV spectral slope) as a function
of rest-frame \lya\ EW for SFGs and their implications on our
understanding of these galaxies.  Results are discussed in
\secref{results}, and we conclude with a summary in \secref{summary}.

Throughout this paper, we assume the standard cosmology with
$\Omega_m$=0.3, $\Omega_{\Lambda}$=0.7 and \Ho=70~km s$^{-1}$
Mpc$^{-1}$.  This corresponds to a look-back time of $\sim$10.4~Gyr at
$z$\,$\simeq$\,2.  Magnitudes are given in the AB system
\citep{oke83}.


\section{Data and Sample Selection}\label{data}

The VUDS observations \citep{lefe15} were done using the
low-resolution multi-slit mode of VIMOS on the VLT. A total of 15
VIMOS pointings ($\sim$224 arcmin$^2$ each, $\sim$1 deg$^2$ total)
were observed with both the LRBLUE and LRRED grisms covering the full
wavelength range from 3650\AA\ to 9350\AA\ in three fields (ECDFS,
VVDS-02h, COSMOS). The total exposure time per pointing was 14h in
each grism. This paper is based on $\sim$80\% of the data, which is
the amount processed at the time of writing this paper. Details of
these observations and the reduction process are described in
\citet{lefe15}.

\subsection{Photometry}

VUDS targeted three well-studied extragalactic fields with 
extensive multi-wavelength photometry. Details are given in 
\citet{lefe15}. Here, we briefly summarize the most relevant imaging
observations for these three fields.

The COSMOS field was observed with HST/ACS in the F814W filter
\citep{scov07,koek07}.  Deep ground based imaging includes
observations in $u^*,B,V,g^+,r^+,i^+,z^+$ bands from Subaru and
CFHT. Details about these and other publicly available
multi-wavelength observations are available at  COSMOS
websites\footnote{ http://cosmos.astro.caltech.edu/data/index.html and
  http://cesam.lam.fr/hstcosmos/}.  The UltraVista survey is acquiring
very deep near-infrared imaging in the $YJHK$ bands using the VIRCAM
camera on the VISTA telescope \citep{mccr12}, while deep Spitzer/IRAC
observations are available through the SCOSMOS \citep{sand07}
program. The CANDELS survey \citep{grog11,koek11} also provided WFC3
NIR photometry in the smaller, central part of the COSMOS field.

The ECDFS field has been the target of many multi-wavelength surveys
in the last decade. The central part of the field, covering
$\sim$\,160~arcmin$^2$, has HST/ACS observations \citep{giav04}
combined with the recent CANDELS HST/WFC3 observations in the
near-infrared bands \citep{koek11}.  The ECDFS field is covered by
deep $UBVRIzJHK$ imaging as part of MUSYC and other surveys in this
field \citep[and references therein]{card10}. The SERVS program
obtained medium-deep observations in 3.6$\mu$m and 4.5$\mu$m
\citep{maud12}, which complement those obtained by the GOODS team
(PI: M. Dickinson) at 3.6, 4.5, 5.6, and 8.0$\mu$m.

The VVDS-02h field has deep optical imaging from CFHT in the
$u^*, g', r', i', z'$ bands \citep[e.g.,][]{cuil12} as part of the
CFHT Legacy Survey. Deep near-infrared imaging in the $JHK_s$ bands
has been obtained with WIRCam on CFHT as part of the WIRCam Deep
Survey \citep[WIRDS;][]{biel12}. The deep near-infrared imaging data
(3.6$\mu$m and 4.5$\mu$m) has been obtained with Spitzer/IRAC as part
of the SERVS program \citep[e.g.,][]{maud12}. We refer the reader to
\citet{lema14a} for a detailed description of multi-wavelength imaging
data in VVDS-02h field.

\subsection{VUDS Target Selection}

The primary selection criterion for galaxies in the VUDS program was
photometric redshifts ($z_{phot}$ + 1$\sigma$\,$\ge$\,2.4), which are
accurate to within 5\% errors for these well-studied fields
\citep[e.g.,][]{ilbe09,coup09,dahl10}.  For high redshift
  ($z$\,$\gtrsim$\,2.5) targets, this photometric redshift selection
  was supplemented by color criteria (e.g., LBG/$ugr$, LBG/$gri$,
  LBG/$riz$), if the objects satisfied these color criteria but were
  not selected from the primary photometric redshift criterion. The
  fraction of targets selected for each criterion at different
  redshifts is shown in Table 2 of \citet{lefe15}. At \ztwo, all
  spectroscopic targets were selected solely based on the photometric
  redshift criterion. Therefore, the targets for the VUDS program
include a representative sample of all star-forming galaxies at a
particular redshift within a given magnitude limit
($i_{AB}$\,$\lesssim$\,25~mag, with some galaxies as faint as
$i_{AB}$\,$\sim$\,27~mag). It is important to note that the target
selection is not based on any particular emission line of a galaxy but
on continuum magnitude.  A detailed discussion about the target
selection, reliability of the redshift measurements and corresponding
quality flags is presented in \citet{lefe15}.

\subsection{Sample Selection}\label{sample}

For this paper, the primary selection criterion for the sample of SFGs
from the VUDS spectroscopic data is the redshift. All objects between
$z$\,=\,2 and $z$\,=\,2.5 are selected in the final sample, keeping
only the best reliability flags (2,3,4,9) --- which gives very high
probability (75-85\%, 95-100\%, 100\%, 80\%, respectively; see
\citealt{lefe15} for details) for these redshifts to be correct. The
total number of galaxies selected at \ztwo\ for the current work is
854.  Based on a simple nearest-neighbor matching to the most recent
catalogs in each field \citep[e.g.,][]{chia05,elvi09,xue11}, we find
that our sample contains minimal contamination from X-ray AGN
($\lesssim$1\%), which suggests that our sample is comprised almost
exclusively of galaxies without powerful X-ray AGN activity. We
  also checked for obscured AGN using \citet{donl12} Spitzer/IRAC
  criteria, and found that $\sim$3\% of galaxies in our sample could
  host AGN based on this criteria.  Some of these IRAC-selected AGN
  are also X-ray AGN, so after removing these overlapping AGN we end
  up with $\sim$2\% of additional AGN candidates using the Donley et
  al. IRAC criteria.  This confirms that our SFGs sample has minimum
  contamination ($\lesssim$3\%) from AGN identified based on their
  X-ray emission and IRAC colors.  This does not rule out the
presence of other types of AGN or AGN-like activity in our sample
\citep[e.g.,][]{cima13,lema10}, but such phenomena do not, typically,
dominate the UV/optical/NIR light of their host galaxies, which is the
important point for our analysis.

  The \texttt{Le PHARE} software package \citep{arno99, ilbe06} was
  used to fit the broad-band observed Spectral Energy Distributions
  (SEDs) with synthetic stellar population models. We use
  BC03 \citep{bruz03} templates to generate a set of stellar population
  models assuming a Chabrier initial mass function \citep[IMF;][]{chab03}, and
  fixed the spectroscopic redshift, while varying metallicity
  ($Z$\,=\,0.4 and 1 $Z_{\odot}$), age (0.05 Gyr
  $\leq$\,$t$\,$\leq$\,$t_{H}$), dust extinction
  (0\,$\leq$\,E$_{\rm s}$(B--V)\,$\leq$\,0.7~mag) --- using a
  \citet{calz00} attenuation law or, in some cases, Calzetti law +
  2175\AA\ bump --- and $e$-folding timescale ($\tau$\,=\,0.1--30~Gyr,
  i.e., from instantaneous burst to continuous star formation) for a
  star-formation history (SFH)\,$\propto$\,exp(-t/$\tau$). We also
  included two delayed SFH models with peaks at 1 and 3 Gyr.  The
  \texttt{Le PHARE} code assumes the \citet{mada95} prescription to
  estimate inter-galactic medium (IGM) opacity.  From the best-fit
  model, we estimate stellar mass, dust extinction E$_{\rm s}$(B--V),
  SFRs, and SSFR for each galaxy. The UV absolute magnitudes are
  measured using the prescription of \citet{ilbe05} and are not
  corrected for the internal dust extinction.

  We emphasize that SFRs and E$_{\rm s}$(B--V)/dust inferred from
    fitting BC03 models to the SED of galaxies give only a rough
    estimate of these physical parameters.  To understand the effect
    of medium-band filters as well as other input quantities on the
    best-fit SED parameters, we compared our SFR and E$_{\rm s}$(B--V)
    measurements with two different surveys that use medium band
    photometry: (1) NEWFIRM Medium Band Survey (NMBS) in the COSMOS
    field \citep{whit12a} and (2) COSMOS SED measurements using 30
    photometric bands \citep{ilbe09}, which includes broad-band as
    well as intermediate and narrow bands in the optical and NIR. For
    NMBS data, we have a small number ($\sim$20) of galaxies matching
    both catalogs. For these galaxies, we find that, on average, SFRs
    differ by $\sim$0.5 dex (factor of $\sim$3), and A$_v$
    (E$_{\rm s}$(B--V)) differ by $\lesssim$\,0.2 ($\lesssim$\,0.07).
    For COSMOS 30-band data which has larger sample ($\sim$300) of
    galaxies, we find that, on average, SFRs differ by $\sim$0.2 dex
    (factor of $\sim$2), and A$_v$ (E$_{\rm s}$(B--V)) differ by
    $\lesssim$\,0.15 ($\lesssim$\,0.05).  It is crucial to note that
    NMBS uses slightly different input set-up for SED-fitting e.g.,
    slightly different SFHs, different choice of extinction laws,
    different metallicity range, use of photometric redshifts,
    compared to what we use, so these choices of input parameters will
    also affect SFRs and dust measurements.  Therefore, the larger
    difference between NMBS and our SED measurements is not only due
    to photometric bands.  As for COSMOS 30-band, we kept all the
    input parameters same as in the broad-band photometry SED fitting
    but changed only the input photometric bands, so other effects do
    not play any role, and we see much smaller differences.  These
    comparisons show that the uncertainties in SFRs and dust
    measurements due to the choice of photometric bands and other
    input SED parameters could be as high as factor of $\sim$3 for
    SFRs and $\sim$0.2 mag for A$_v$ (or $\sim$0.07 for
    E$_{\rm s}$(B--V)), which is typically expected for the
    SED-fitting process \citep[e.g.,][]{muzz10,most12}. It is
    important to mention that these uncertainties are inherent in the
    SED fitting process and are not unique to this study. Moreover,
    this study is based on the relative measurements of three
    different galaxy populations and as long as there are no drastic
    differences in the IMF or SFHs of these populations, these
    systematic uncertainties does not play a significant role in the
    conclusions of this paper.

The contribution of major emission lines in different filters has been
included in the models of the \texttt{Le PHARE} code. We use these
templates with emission lines to carry out SED fitting.  Neglecting
emission lines during the SED fitting process can overestimate the
best-fit stellar ages and stellar masses by up to $\sim$\,0.3 dex
\citep[e.g.,][]{scha09,fink11,atek11}. The \texttt{Le PHARE} code
accounts for the contribution of emission lines with a simple recipe
based on the \citet{kenn98} relations between the SFR and UV
luminosity, H$\alpha$ and [OII] lines. The code includes the \lya,
H$\alpha$, H$\beta$, [OII], OIII[4959] and OIII[5007] lines with
different line ratios with respect to the [OII] line, as described in
\citet{ilbe09}.

\subsection{Sample Properties}

\begin{figure*}
   \centering
    \includegraphics[scale=0.55]{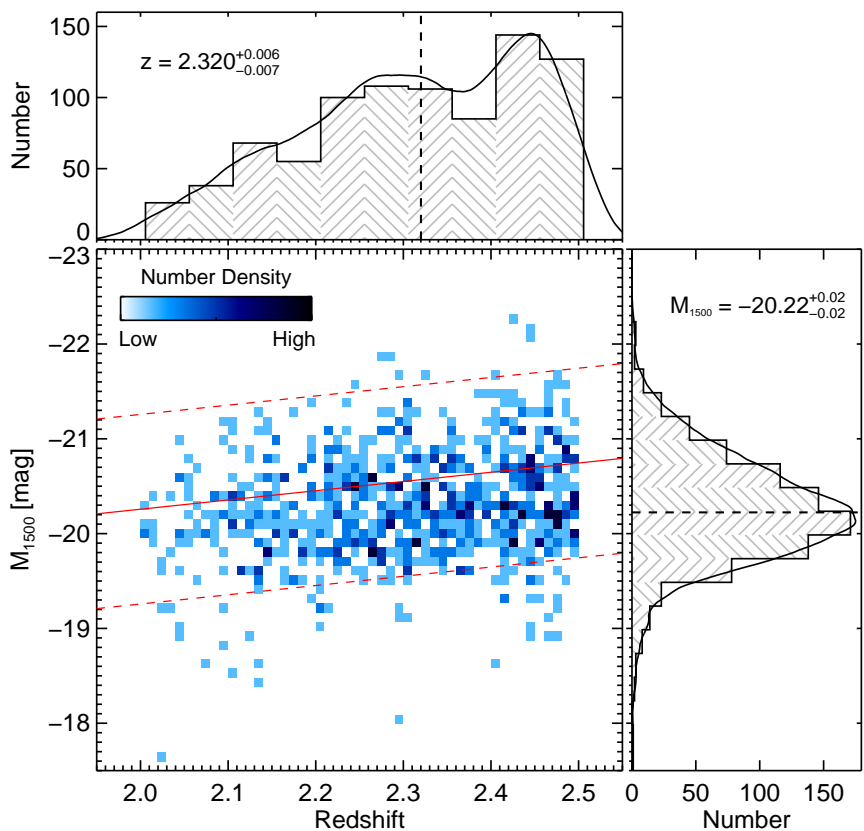}

    \caption{Redshift versus UV absolute magnitude (M$_{\rm 1500}$)
      for spectroscopically confirmed SFGs at \ztwo\ from VUDS. The
      median M$_{\rm 1500}$ of the sample is --20.22 mag, the median
      redshift is 2.320. The quoted uncertainties are the errors on
      the median ($1.253\,\times\,\sigma/\sqrt{\rm N_{\rm gal}}$) estimated
      from the 1$\sigma$ dispersion of the distribution and the total
      number of galaxies, where $\pm$\,1$\sigma$ corresponds to the
      range between the 16th and the 84th percentile values of the
      distribution. The red solid line indicates the evolving M$^*$
      (characteristic magnitude) as a function of redshift as derived
      from the compilation of \citet{hath10}. The dashed red lines
      indicate M$^* \pm$1 mag. The black solid curve on the histogram
      is the kernel density estimation (KDE) of the distribution,
      which is a smoother non-parametric density estimator compared to
      the histogram and is not affected by the bin-size or the end
      points of the bins. The density of points is color-coded as
      shown in the color-bar.}
         \label{fig:z_mag}
   \end{figure*}

\figref{fig:z_mag} shows distributions of spectroscopic redshifts and
far-UV (1500\AA) absolute magnitudes (M$_{\rm 1500}$) for the full SFG
sample at \ztwo.  The median M$_{\rm 1500}$ of the sample is --20.22 mag,
while the median redshift is 2.320. The figure also shows the evolving
characteristic magnitude M$^*$ as a function of the redshift, as
derived from the compilation of \citet{hath10}. The M$^*\pm$1~mag
range covers most of our data which implies that we are probing
luminosities around L$^*$ at these redshifts. The shape of the
redshift distribution is dictated, in part, by the selection function
used to select target galaxies for VUDS. The main selection criteria
under which primary targets are selected, is $z_{phot}$ +
1$\sigma$\,$\ge$\,2.4. Therefore, at lower redshifts
(2\,$<$\,$z$\,$<$\,2.4) we are on the tail of the probability
distribution of photometric redshifts, and this explains the
decreasing number of galaxies at $z$\,$<$\,2.4. This bias equally
affects all galaxies --- galaxies with and without \lya\ in emission
--- so the general results in this paper, making relative comparisons
of the two populations, are not affected by this selection bias. The
second reason for the small number of galaxies close to the lower
limit of our redshift range could be the sensitivity of the VIMOS
instrument which starts to decrease dramatically below 3800\AA. This
means that we might not be able to properly identify continuum breaks,
important features to secure the redshift, below redshift
$z$\,$\lesssim$\,2.15. In this redshift range, we might be biased in
favor of strong \lya\ emitters as we could still identify those even
at $z$\,$\lesssim$\,2.15.  To check this sample selection effect, we
cut our sample at redshift $z$\,=\,2.15. By doing so, we remove
$\sim$10\% of galaxies from the full sample, which is very small
compared to the large sample size of VUDS galaxies.  This small change
in number of galaxies does not change our final results.  Therefore,
the results presented in this paper are not affected by this selection
bias.
%
%
%

\begin{figure*}
   \centering
   \includegraphics[scale=0.55]{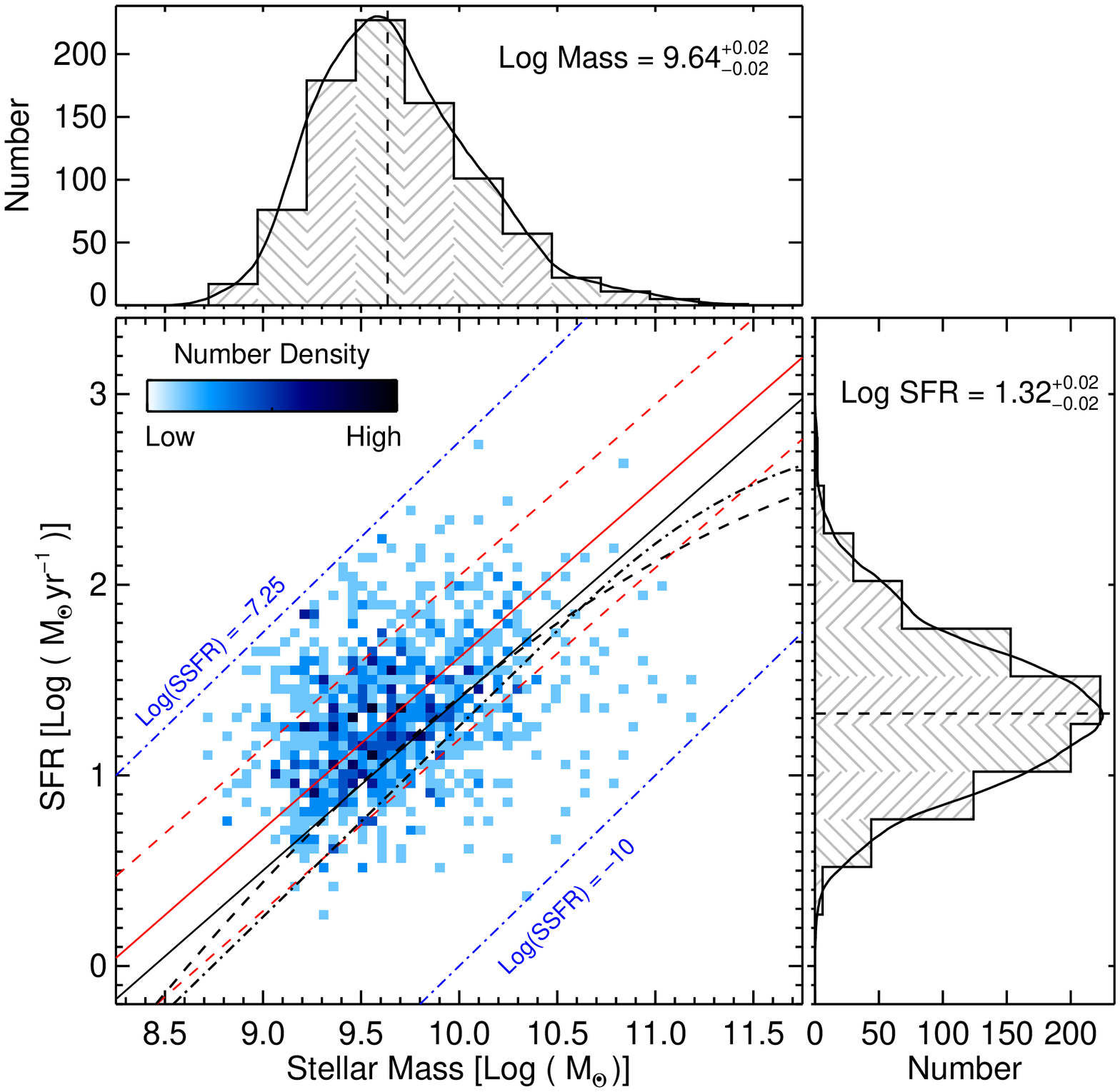}

   \caption{SFR versus stellar mass for SFGs at \ztwo\ in VUDS. The
     full sample has median stellar mass of
     10$^{9.64}$\,$\sim$\,4$\times$10$^{9}$ M$_{\odot}$ and median SFR
     of 10$^{1.32}$\,$\sim$\,20 M$_{\odot}$ per year. The quoted
       uncertainties are the errors on the median
       ($1.253\,\times\,\sigma/\sqrt{\rm N_{\rm gal}}$) estimated from
       the 1$\sigma$ dispersion of the distribution and the total
     number of galaxies, where $\pm$1$\sigma$ corresponds to the range
     between the 16th and the 84th percentile values of the
     distribution.  The black solid curve is the KDE of the
     distribution.  The black solid line shows the main sequence of
     SFGs as defined by \citet{dadd07} at $z$\,$\sim$\,2, and the
     solid red line is the best-fit relation (with the slope fixed to
     the value from Daddi et al.) for our sample with the total
     scatter ($\sim$0.4 dex) shown by red dashed lines. The black
       dashed line is the main sequence relation from \citet{whit14},
       while the black dot-dashed line is the relation from
       \citet{schr15}. The blue dot-dashed lines show limits on SSFR
     for these galaxies. The density of points is color-coded as shown
     in the color-bar. }

         \label{fig:mass_sfr}
   \end{figure*}

   \figref{fig:mass_sfr} shows distributions of stellar mass and SFR
   for the full SFG sample at \ztwo. The median stellar mass of the
   sample is $\sim$4$\times$10$^{9}$ M$_{\odot}$ and the median SFR is
   $\sim$20 M$_{\odot}$ per year. We show the `main sequence' of SFGs
   as defined by \citet{dadd07} at $z\simeq2$, and the best-fit
   relation for our sample with the corresponding scatter ($\sim$0.4
   dex).  We keep the logarithmic slope fixed to the value from Daddi
   et al. relation because the redshift probed is very similar for
   both studies.  If we fit the observed SFR-stellar mass relation
   without keeping the slope fixed, then we get the logarithmic slope
   of $\sim$0.66 which is flatter than the slope of 0.90 obtained by
   Daddi et al. at similar redshifts.  \figref{fig:mass_sfr} shows
   that our sample has a large scatter in the stellar mass-SFR plane
   covering a large range in SFR and over two orders of magnitude in
   stellar mass.  It is important to note that our best-fit relation
   (keeping the logarithmic slope fixed at 0.90) between SFR and
   stellar mass is $\sim$0.2 dex off from the Daddi's SF main
   sequence.   The Daddi et al. sample is at slightly lower
     redshift compared to our sample, so we also show in
     \figref{fig:mass_sfr} the best-fit main sequence relation from
     \citet{whit14} and \citet{schr15}. The redshift probed by these
     studies is very similar to ours. We still find a similar offset
     ($\sim$0.2 dex) between our main sequence relation and the
     relation from these studies, which predicts slightly lower
     normalization i.e., lower SFR at a given stellar mass.  This
     difference could be due to the fact that we estimate SFRs based
     on the SED fitting process which has wide variety of SFHs and it
     gives slightly different SFR compared to SFR(UV) or SFR(UV+IR).
     When we estimate SFR using UV luminosity (corrected for dust
     using the UV slope), we estimate average SFR to be $\sim$0.2 dex
     lower compared to SFR(SED). Therefore, this could be a reason for
     $\sim$0.2 dex offset we see in the main sequence relation
     compared to other studies, as they use SFR(UV) or SFR(UV+IR) in
     their relations.  By selection, and confirmed by $NUV-r-J$ plot,
     our galaxies are not very dusty and hence, not many galaxies have
     IR/FIR detections. In future papers, we will investigate IR SFRs
     for these galaxies based on their 24$\mu$m and Herschel
     observations.

   The scatter (total scatter obtained from the fitting process) in
   the SFR-stellar mass correlation is $\sim$0.4 dex, which is higher
   than that of \citet{dadd07} by $\sim$0.2 dex.  We believe that
     this large scatter is caused by the choice of wide range of SFHs
     in the SED fitting process. Various studies
     \citep[e.g.,][]{mclu11,scha13} have shown that a broader range in
     SFHs in the SED fitting process could have strong implications on
     the scatter of the main sequence relation.  Considering that our
     sample has similar redshift ($z$\,$\simeq$\,2) as Daddi et
     al. and other studies, these differences in the main sequence
     relation emphasize that the sample properties as well as
     measurement techniques play a significant role in the
     determination of the correlation between SFR and stellar masses
     of galaxies \citep[see, e.g.,][for uncertainties in SED fitting
     process]{pfor12}. To underscore the importances of these
     differences, we point out few subtleties between the Daddi et
     al. relation and ours. Our sample was selected based on the
   photometric redshift technique, while the \citet{dadd07} sample was
   selected based on a single $BzK$ color technique, highlighting that
   sample selection likely plays a significant role in determining the
   parameters of the stellar mass-SFR relation. We use different
   star-formation histories in our SED-fitting technique that
   estimates SFR and stellar masses, while \citet{dadd07} obtain SFRs
   directly from UV luminosity and their stellar masses are based on a
   single color technique \citep{dadd04}. This difference in method
   also likely influences the measurements of these stellar parameters
   and thus the shape \& normalization of the stellar mass-SFR
   relation. The detailed discussion about various aspects of this
   relation is beyond the scope of this paper, therefore, for in-depth
   study on this relation and its evolution with redshift, we refer
   the reader to \citet{tasc15}, and for more details on the effect of
   different SFHs on the SFR-stellar mass correlation, we refer the
   reader to Cassar\`a~et~al. (in~prep). Both these investigations are
   based on the same VUDS dataset, but over a larger redshift range.
%
%
%

\section{UV Spectral Slope ($\beta$)}\label{uvslope}

\begin{figure*}
   \centering
   {\includegraphics[scale=0.45, angle=0]{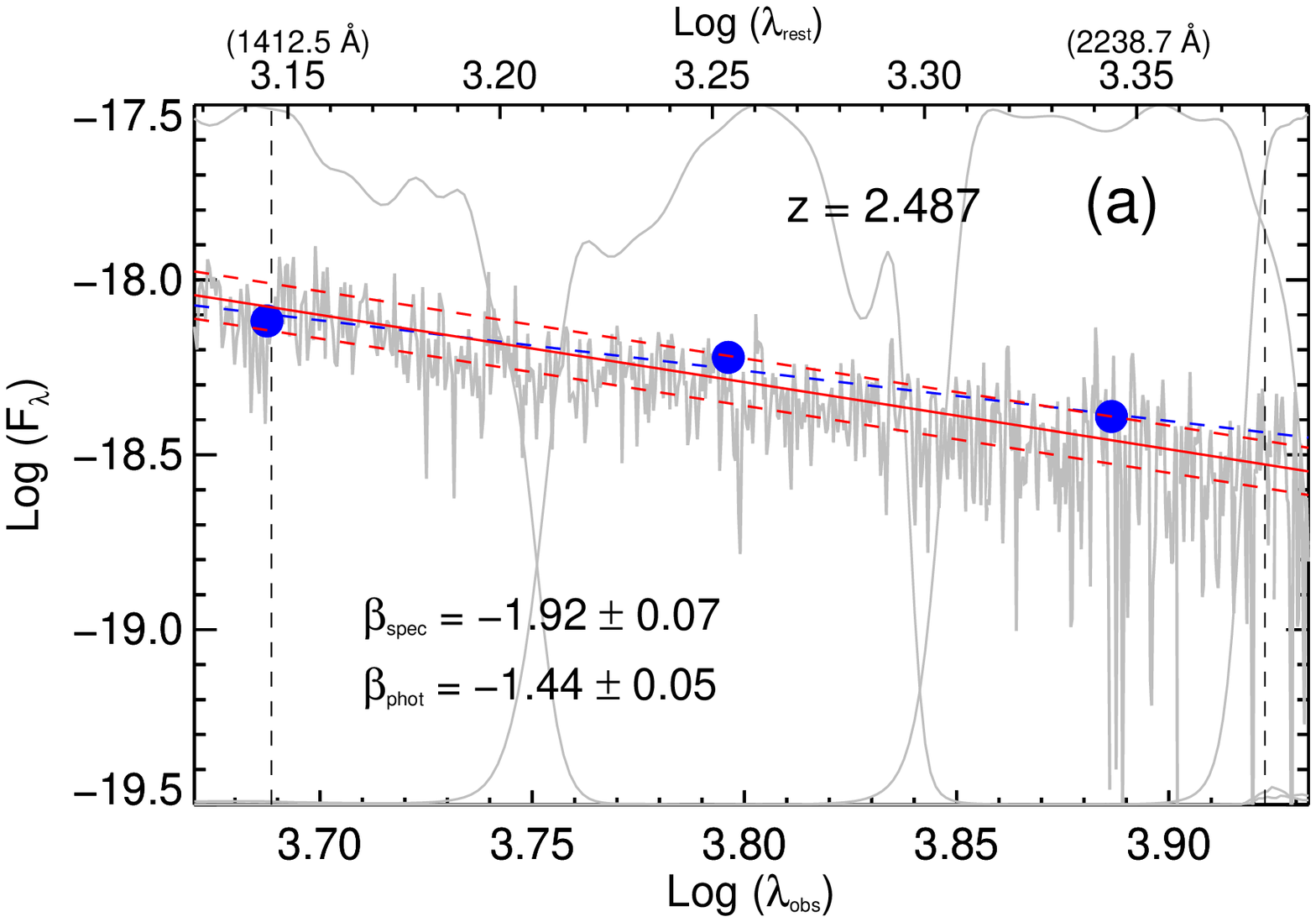}
   \includegraphics[scale=0.45, angle=0]{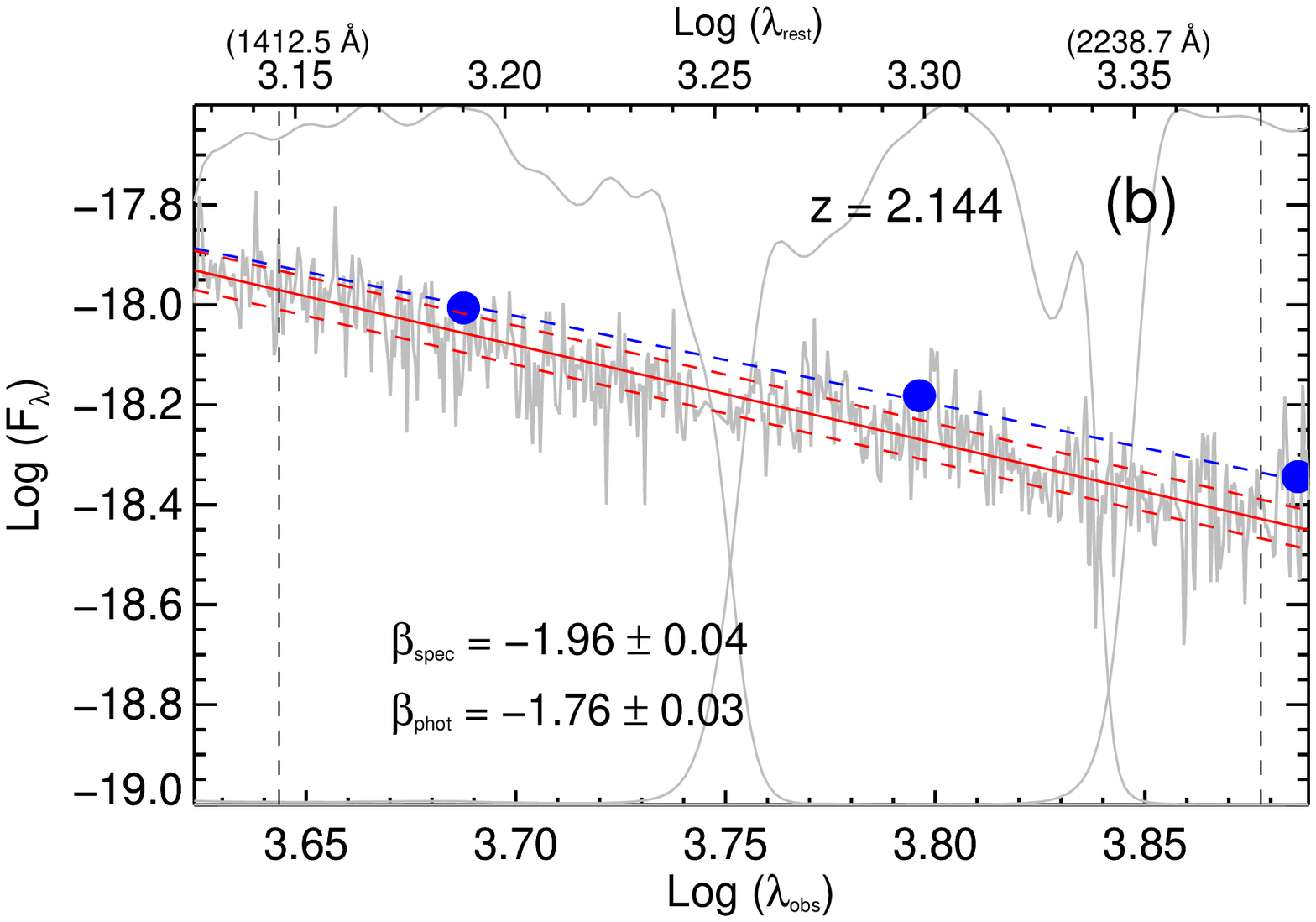}
   \includegraphics[scale=0.45, angle=0]{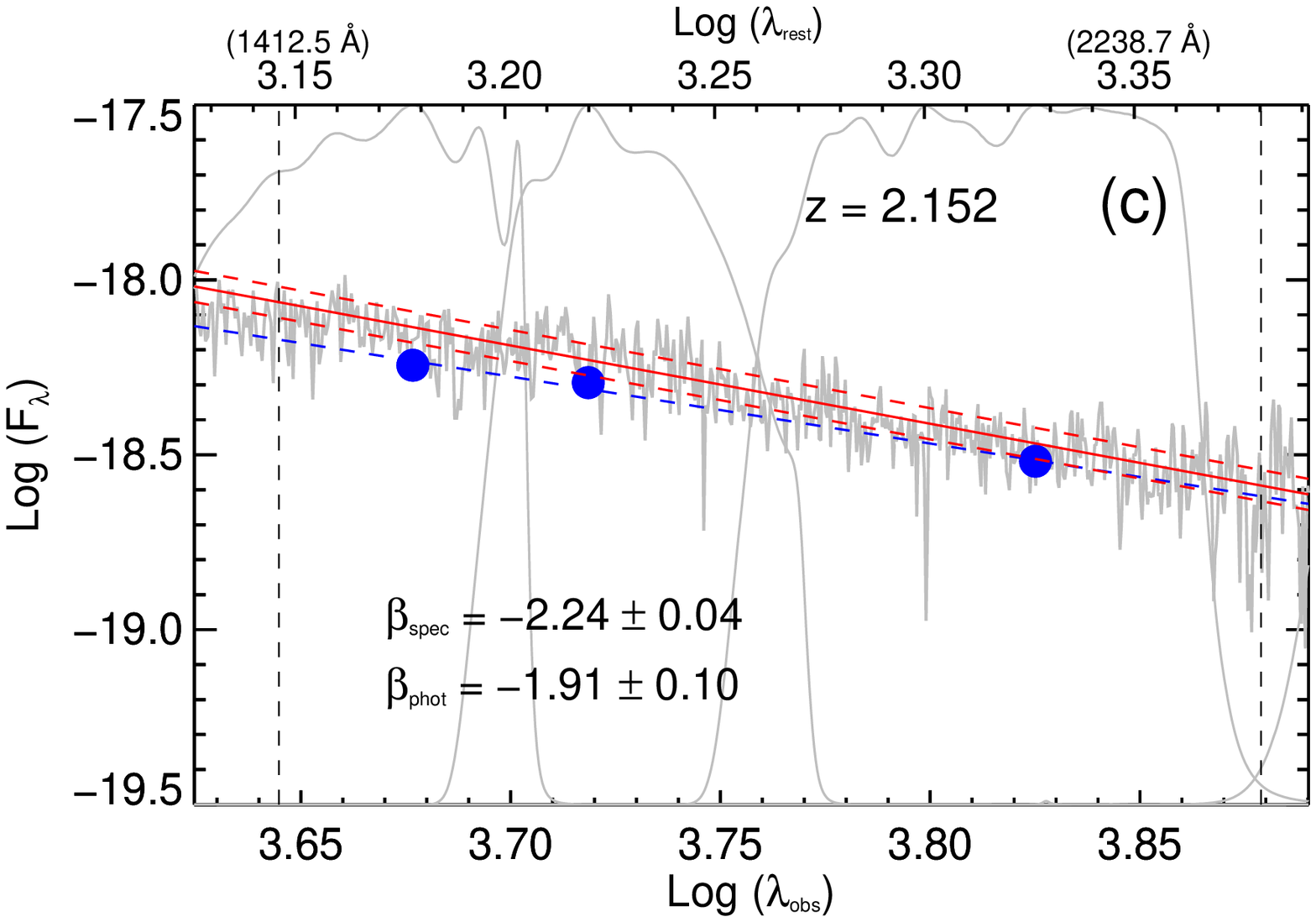}
   \includegraphics[scale=0.45, angle=0]{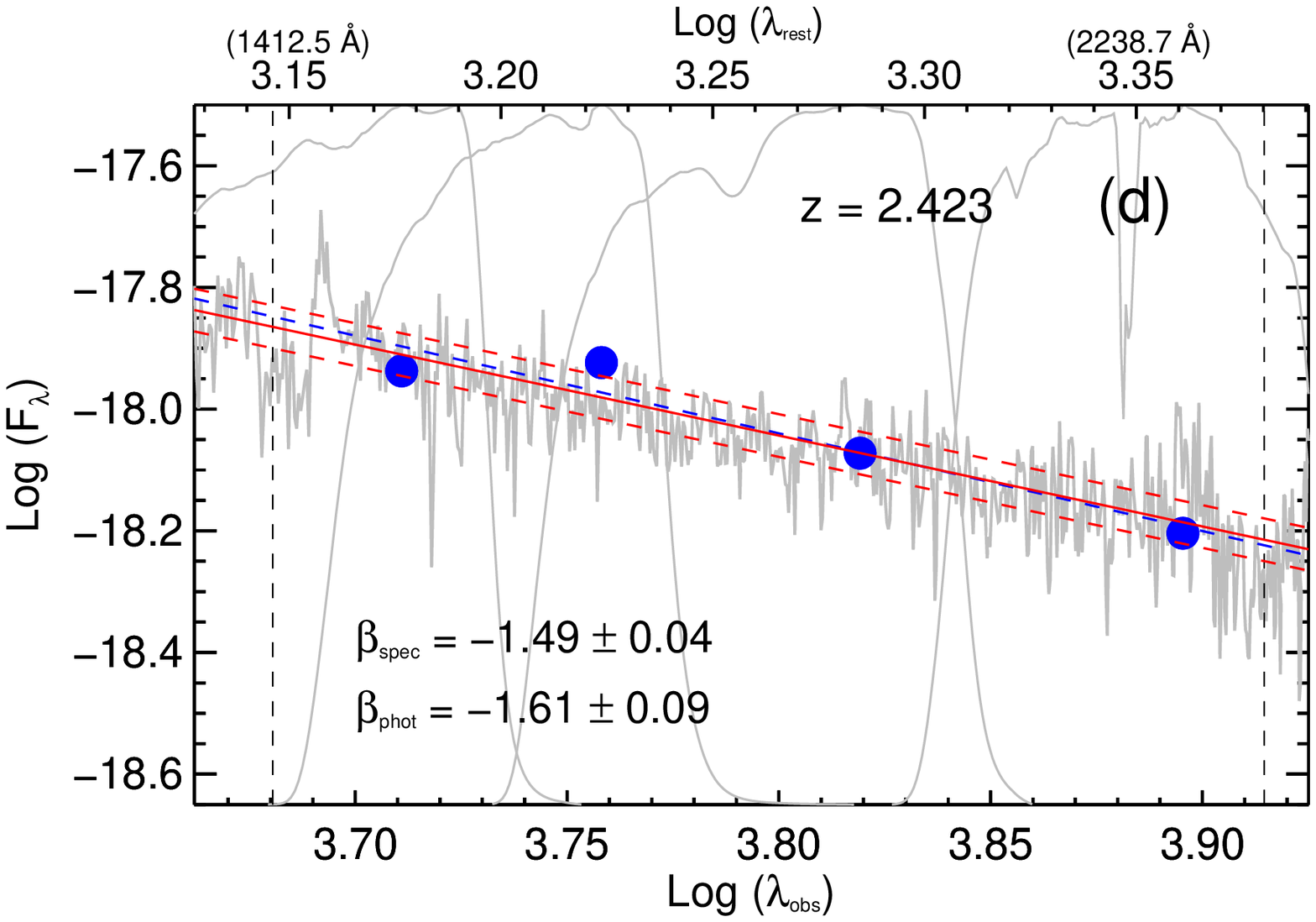}}

 \caption{Examples of UV spectral slope ($\beta$) fitting on the VUDS
   spectra (shown in grey).  The spectra between rest-frame 1400\AA\
   to 2400\AA\ (between dashed vertical lines) are fitted with a
   linear relation (solid red lines) between Log($\lambda$) vs
   Log(F$_{\lambda}$) after masking out known lines and removing
   spurious noise spikes. The dashed red lines are 1$\sigma$
   deviations in the best-fit relation. The blue filled circles are
   the photometric magnitudes from the broad-band filters (grey
   curves) used to measure the photometric $\beta$, while the blue
   dashed line is a linear fit between these photometric data points
   to estimate the photometric UV slope.}
         \label{fig:beta_measure}
   \end{figure*}

The UV spectral slope ($\beta$) is determined from a power-law fit to
the UV continuum spectrum \citep{calz94},
$f_{\lambda}$\,$\varpropto$\,$\lambda^{\beta}$, where $f_{\lambda}$ is
the flux density per unit wavelength (ergs s$^{-1}$ cm$^{-2}$
\AA$^{-1}$). We use the VUDS spectra of SFGs at \ztwo\ to estimate
their UV slope by fitting a linear relation to the full wavelength
range between rest-frame 1400 and 2400\AA\ in the observed
(Log($\lambda$) versus Log( $f_{\lambda}$)) spectrum.  The rest-frame
wavelength range (1400--2400\AA) includes 7 out of 10 spectral fitting
windows identified by \citet{calz94} to estimate the UV spectral slope
for local starburst galaxies, and takes full advantage of the spectral
coverage of the VUDS spectra at these
redshifts. \figref{fig:beta_measure} shows four examples of UV
spectral slope fitting on the VUDS spectra. Before fitting, the
spectra were cleaned by masking out known lines and removing any
spurious noise spikes.  The fitting was done using the IDL routine
LINFIT, which fits the data to the linear model by minimizing the
$\chi^2$ statistic. The running dispersion values measured on the
clean spectra are used as measured errors, and the routine uses these
errors to compute the covariance matrix. In \figref{fig:beta_measure},
the solid red line is the best-fit spectroscopic slope while dashed
red lines show 1$\sigma$ uncertainties associated with the best-fit
relation.  The formal statistical uncertainty in the best-fit
  $\beta_{\rm spec}$, which is measured over a large number of points,
  is smaller than the uncertainty in photometric $\beta$ measured
  using few photometric points.  For comparison, we also measure the
UV spectral slope directly from the broad-band photometry which
encompasses a similar rest-frame wavelength range.  We used $BVRI$
  bands for the ECDFS field, $g^+Vr^+i^+$ for the COSMOS field, and
  $g'r'i'z'$ bands for the VVDS field. These photometric bands cover a
  rest-frame wavelength range from $\sim$1200\AA\ to $\sim$3000\AA,
  depending on the redshift. \figref{fig:beta_measure} also shows
the photometric magnitudes of the broad-band filters which are used to
measure the photometric UV slope, and the corresponding best-fit
photometric UV slope.  \figref{fig:beta_hist} shows the UV spectral
slope ($\beta$) distribution as measured from the VUDS spectra
($\beta_{\rm spec}$) and the broad-band photometry
($\beta_{\rm phot}$).  The median $\beta_{\rm spec}$ is
  --1.36\,$\pm$\,0.02, and the median $\beta_{\rm phot}$ is
  --1.32\,$\pm$\,0.02. The quoted uncertainties are the errors on the
  median ($1.253\,\times\,\sigma/\sqrt{\rm N_{\rm gal}}$).  It is
worth noting that even if we use two or three photometric bands,
covering the similar wavelength range as four bands, we get similar
median values for the photometric UV slope of the whole sample, 
  which is in the range of --1.40$\pm$0.10.  We also measured UV
  slopes by convolving the spectra with the photometric filter curves,
  and found that the median value is --1.31\,$\pm$\,0.03, which is
  very similar to the median photometric and spectroscopic UV slopes.

%
%
   In \figref{fig:beta_hist} , and in subsequent figures of this
   paper, we will calculate statistical significance for the
   distributions and correlations in the following two ways: (a) for
   histograms, we use KS-statistics on the distributions. The
   probability of KS-statistics (P$_{\rm KS}$) measures whether two
   sets of data are drawn from the same parent distribution or
   not. Small values (close to zero) of P$_{\rm KS}$ show that the
   cumulative distribution functions of the two sets of data are
   significantly different. This is measured using the IDL routine
   KSTWO, and (b) for correlations, we use the Spearman correlation
   test.  Values of the Spearman correlation coefficient (\rsc) close
   to $\pm$1 imply strong monotonic correlation, while values close to
   0 imply no significant correlation between two variables. The
   significance level is denoted by P$_{\rm SC}$, where small values
   of P$_{\rm SC}$ imply significant correlation.  It is important to
   note that for a large sample size, very small differences (or small
   \rsc\ values) in two sets of data will be detected as highly
   significant. When a statistic is significant, it simply means that
   the difference is real. It does not mean the difference is large or
   important.  The Spearman correlation coefficient is measured using
   the IDL routine R\_CORRELATE, and is measured from all galaxies as
   well as from the median-binned values.This routine can also give
   the value of the sum-squared difference of ranks, and the number of
   standard deviations ($\sigma$) by which it deviates from its
   null-hypothesis expected value.

   In \figref{fig:beta_hist} (left panel), the median values of
     $\beta_{\rm spec}$ and $\beta_{\rm phot}$ are very similar and
     differ by only $\sim$0.04.  This difference is not 
     significant considering that the errors on the median values is
     $\pm$0.02. However, the full distributions of $\beta_{\rm spec}$ and
     $\beta_{\rm phot}$ are marginally different as confirmed by
     P$_{\rm KS}$\,$\sim$\,0.00057, which implies that a null
     hypothesis (similar distributions) is rejected at $\sim$3$\sigma$
     level.  To investigate the difference between individual
     $\beta_{\rm spec}$ and $\beta_{\rm phot}$, we look at four
     examples shown in \figref{fig:beta_measure}.
     \figref{fig:beta_measure}(a) shows an extreme example in which
     the difference between two $\beta$s is $\sim$0.5, though
     $\beta_{\rm phot}$ is within the 1$\sigma$ uncertainties of
     $\beta_{\rm spec}$.  \figref{fig:beta_measure}(b) and
     \figref{fig:beta_measure}(c) show examples where
     $\beta_{\rm phot}$ is just outside the 1$\sigma$ uncertainties in
     $\beta_{\rm spec}$ and we find that both these $\beta$ values
     differ by $\sim$0.2-0.3.  \figref{fig:beta_measure}(d) shows an
     example where the difference between the two $\beta$s is very
     small ($\sim$0.1) and well within the measurement uncertainties.

   The middle panel of \figref{fig:beta_hist} shows the direct
     comparison between individual $\beta_{\rm spec}$ and
     $\beta_{\rm phot}$, while the bottom plot in the middle panel
     shows the difference between $\beta_{\rm spec}$ and
     $\beta_{\rm phot}$ as a function of $\beta_{\rm spec}$.  The best
     orthogonal fit relation between $\beta_{\rm spec}$ and
     $\beta_{\rm phot}$ is flatter than 1-to-1 relation, and is given
     by $\beta_{\rm phot}=-0.43 + 0.63 \times \beta_{\rm spec}$. The
     total scatter in the $\beta_{\rm spec}$ vs $\beta_{\rm phot}$
     relation could be as high as $\sim$0.5.  The uncertainties in
     both $\beta$ measurements could be the main cause of the flatter
     slope ($<$\,1) and the large scatter.  The average measurement
     error in $\beta_{\rm spec}$ is $\sim$0.08, while it is $\sim$0.15
     for $\beta_{\rm phot}$.  Depending on the object and its
     redshift, it is possible to have additional factors affecting
     $\beta$ measurements of individual galaxies, as discussed below.

     For $\beta_{\rm spec}$, there are two potential sources of
     uncertainties. First, the bluest part of the spectra is sensitive
     to various atmospheric and galactic corrections made to the
     spectra \citep[see][for details]{lefe15,thom14}, and even though
     these corrections are carefully applied to minimize the flux
     loss, there are uncertainties in the bluest part of the
     spectra. The UV slope measurement is sensitive to the changes in
     the anchor blue region of the spectra and hence, this could
     affect the UV slope measurements for some objects. Second, as
     seen in \figref{fig:beta_measure}(a), the spectroscopic $\beta$
     is fitted in such a way that it avoids the more noisy red part of
     the spectrum to minimize its effect on the best-fit slope, which
     could make $\beta_{\rm spec}$ steeper or bluer compared to
     $\beta_{\rm phot}$. While this fitting effect is kept at the
     minimum level, any small difference in the fitting of the red
     part of the spectrum could lead to slightly bluer
     $\beta_{\rm spec}$ compared to $\beta_{\rm phot}$. These
     uncertainties in $\beta_{\rm spec}$ are estimated to be small
     ($\lesssim$0.1) but could vary from object to object.

     For $\beta_{\rm phot}$, the wavelength range over which $\beta$
     is measured has a great influence on the resulting $\beta$
     values. \citet{calz94} defined a wavelength baseline (1250\AA\ to
     2600\AA) to measure $\beta$ for galaxies at lower redshifts and
     this range is widely used for all measurements.  But as discussed
     in \citet{calz01}, the UV slopes fitted with the longer
     wavelength baseline are always bluer (more negative) compared to
     the UV slopes from shorter baseline \citep[see
     also][]{meur99,tali15}.  The main reason is that the slopes
     fitted with the longer wavelength baseline cover the
     2300-2800\AA\ range which has a large decrement due to a large
     number of closely spaced FeII absorption lines and this low
     continuum gives them steeper/bluer slopes. Our $\beta_{\rm phot}$
     is measured using broad photometric bands, and depending on the
     exact redshift of a galaxy the wavelength range covered by these
     bands could be as large as $\sim$1200\AA\ to $\sim$3000\AA. This
     range includes the wavelength range (1400\AA\ to 2400\AA) used to
     measure $\beta_{\rm spec}$, but $\beta_{\rm phot}$ has a longer
     wavelength baseline for many galaxies which could lead to bluer
     $\beta_{\rm phot}$ for these galaxies.  The other factor
     affecting the $\beta_{\rm phot}$ measurement is the contribution
     of strong \lya\ emission to the flux of the bluest band used for
     measuring $\beta_{\rm phot}$.  For a small number of strong \lya\
     emitters, it is likely that the bluest band used in the
     $\beta_{\rm phot}$ measurement is affected by the strong \lya\
     emission line.  To check this effect, we measured $\beta$s for
     galaxies with no \lya\ emission and for galaxies with strong
     \lya\ emission. We find that the difference between the median
     values of $\beta_{\rm phot}$ and $\beta_{\rm spec}$ is smaller
     ($\beta_{\rm spec}$ bluer by $\sim$0.05) for galaxies with no
     \lya\ emission compared to galaxies with strong \lya\ emission,
     where on average $\beta_{\rm phot}$ is bluer compared to
     $\beta_{\rm spec}$ by $\sim$0.13.  Hence, strong \lya\ line in
     the bluest photometric band could lead to higher flux in that
     band which in turn could give steeper/bluer $\beta_{\rm phot}$.
     This is confirmed by our estimation of Ly$\alpha$ contamination
     to the bluest photometric band by comparing flux densities with
     and without \lya\ line in the broadband. We found that the
     contamination in the broadband for the strongest Ly$\alpha$
     emission lines could be $\sim$0.05--0.1 mag. This difference in
     the magnitude of bluest band could cause $\sim$0.1--0.2
     difference in the photometrically measured UV slope. These
     uncertainties in $\beta_{\rm phot}$ are estimated to be
     $\lesssim$0.2, but could vary from object to object. 

   In summary, the average $\beta_{\rm spec}$ and
     $\beta_{\rm phot}$ are very similar for the VUDS SFG sample, and
     the uncertainties discussed above could affect individual
     measurements, as shown in the middle panel of
     \figref{fig:beta_hist}. The detailed comparison between
     individual $\beta$s, and quantifying various sources of
     uncertainties for such a large sample, requires an in-depth
     study, which is beyond the scope of this paper.

   The median $\beta_{\rm spec}$ value of the whole SFG sample, 
     --1.36$\pm$0.02  with 1$\sigma$ scatter of $\sim$0.5,
   is consistent with the UV slope estimates obtained at similar
   redshifts using ground-based spectra. \citet{tali12} obtained
   $\beta_{\rm spec}$ of --1.11$\pm$0.44 (rms) for 74 SFGs at
   $z$\,$\simeq$\,2 using VLT/FORS2 spectra, while \citet{noll05} used
   FORS spectroscopic data of 34 UV-luminous galaxies at \ztwo\ and
   found $\beta_{\rm spec}$ for the sub-sample of galaxies to be
   --1.01$\pm$0.11. Both these $\beta$ values are roughly consistent
   with our median $\beta_{\rm spec}$ when considering the additional
   uncertainty introduced by the fact that both studies used slightly
   different wavelength range ($\sim$1200-2600\AA, and
   $\sim$1200-1800\AA, respectively), from each other \& our own, to
   fit the UV slope.  In addition, the median photometric UV slope
   obtained here, --1.32$\pm$0.02 with  1$\sigma$
     scatter of $\sim$0.5, is very similar to the average UV slopes
   measured at these redshifts (\ztwo) from HST photometry (e.g.,
   --1.58$\pm$0.27 (1$\sigma$) from \citealt{bouw09} and
   --1.71$\pm$0.34 (1$\sigma$) from \citealt{hath13}).  These $\beta$
   values --- spectroscopic and photometric --- are also in good
   agreement with the evolution of $\beta$ with redshift which shows
   that the average $\beta$ at \ztwo\ is redder compared to $\beta$ at
   higher redshifts \citep[e.g.,][]{hath08b, bouw09, fink12}.

\subsection{$\beta$ versus E$_{\rm s}$(B--V)}

\begin{figure*}
   \centering
   \includegraphics[scale=0.5, angle=0]{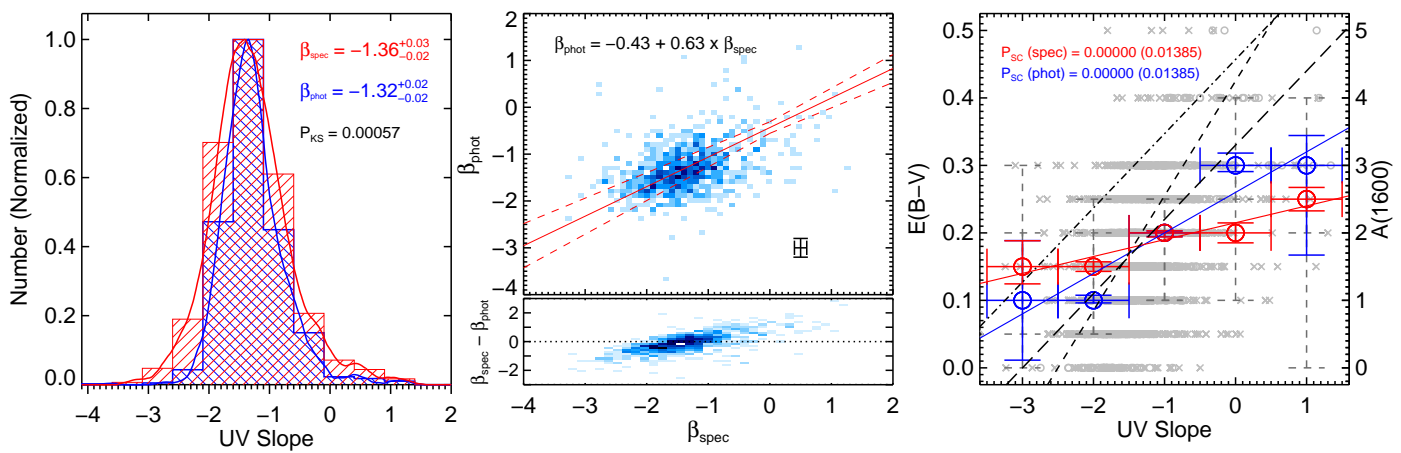}

   \caption{(Left) UV spectral slope ($\beta$) distribution as
     measured from the VUDS spectra (red histogram) and the broad-band
     photometry (blue histogram). The solid curve is the KDE of the
     distribution. The median $\beta$ values as well as KS-test
     probability (P$_{\rm KS}$) show very small difference in these
     $\beta$s.  (Middle) $\beta_{\rm spec}$ versus
       $\beta_{\rm phot}$.  The solid red line is the best orthogonal
       fit relation between these two $\beta$s, while the dashed lines
       show the uncertainties in the best-fit relation. This figure
     also shows the difference ($\beta_{\rm spec}$-$\beta_{\rm phot}$)
     as a function of $\beta_{\rm spec}$. The dotted line is for
     $\beta_{\rm spec}$-$\beta_{\rm phot}$\,=\,0. The density of
     points is color-coded as in previous figures.  Average
       uncertainties in these measurements are plotted in the
       bottom-right corner. (Right) The correlation between the UV
     slope and the SED derived E$_{\rm s}$(B--V) is such that the
     redder UV slope implies dustier [higher E$_{\rm s}$(B--V)]
     galaxies. The error bars in \emph{x} illustrate the sizes of the
     bins, while the errors in \emph{y} are $\pm$1$\sigma$ scatter
     (dashed error bars) corresponding to the range between the 16th
     and the 84th percentile values within each bin, while the smaller
     solid error bars are the errors on the median values
     ($1.253\,\times\,\sigma/\sqrt{\rm N_{\rm gal}}$).  The red
     circles are for the spectroscopic $\beta$, while the blue circles
     are for the photometric $\beta$. The black small-dashed line
       is the dust-UV slope relation at
       1\,$\lesssim$\,$z$\,$\lesssim$\,2 from \citet{buat12}, the
       black dot-dashed line is the dust-UV slope relation at
       $z$\,$\sim$\,2 from \citet{oteo14}, and the black long-dashed
       line is the dust-UV slope relation at 1\,$<$\,$z$\,$<$\,3 from
       \citet{tali15}.  The background grey points are the values for
     all galaxies (crosses are $\beta_{\rm spec}$ and open circles are
     $\beta_{\rm phot}$). The statistical significance from the
     Spearman correlation coefficient is shown to confirm the
     correlation. The corresponding A$_{1600}$ (shown on the right
     y-axis) is estimated from E$_{\rm s}$(B--V) using the Calzetti
     law
     \citep[A$_{1600}$\,$\sim$\,10$\times$E$_{\rm
       s}$(B--V);][]{calz00}.}

         \label{fig:beta_hist}
   \end{figure*}

The UV radiation is absorbed by dust and heated dust particles re-emit
absorbed energy in the mid- and far-infrared. The ratio of
L$_{\rm IR}$-to-L$_{\rm UV}$ is the direct measure of dust attenuation
in a galaxy. A higher L$_{\rm IR}$/L$_{\rm UV}$ ratio implies a larger
dust content. There is an excellent correlation between
L$_{\rm IR}$/L$_{\rm UV}$ and the UV spectral slope at $z$\,$\sim$\,2
\citep[e.g.,][]{seib02,redd12}, which is in good agreement with the
local measurements \citep[e.g.,][]{meur97,meur99}, in that, an
increase in L$_{\rm IR}$/L$_{\rm UV}$ corresponds to redder UV
slopes. For SFGs at \ztwo\ in the VUDS sample, the correlation between
the UV slope and SED derived E$_{\rm s}$(B--V) is shown in
\figref{fig:beta_hist} (right panel). A positive correlation is
observed between $\beta$ and E$_{\rm s}$(B--V) implying redder UV
slope for dustier (higher E$_{\rm s}$(B--V)) galaxies. We show the
correlation of E$_{\rm s}$(B--V) with both, photometric and
spectroscopic, $\beta$s. The statistical significance of this
correlation is measured by the Spearman correlation test.  The
Spearman correlation coefficient (\rsc) is 0.25 (0.95) for all
(binned) $\beta_{\rm spec}$ values and 0.65 (1.00) for all (binned)
$\beta_{\rm phot}$ values, which implies high significance
(P$_{\rm SC}$\,$\sim$\,0) for a strong positive correlation. The
significance of \rsc\ as measured by the number of standard deviations
by which the sum-squared difference of ranks deviates from its
null-hypothesis expected value is $\sim$7$\sigma$ for all
$\beta_{\rm spec}$ values and $\sim$18$\sigma$ for all
$\beta_{\rm phot}$ values.  The correlation between $\beta_{\rm phot}$
and E$_{\rm s}$(B--V) is stronger, in part, because they are both
estimated from the same photometry. The $\beta_{\rm spec}$ correlation
is flatter compared to the $\beta_{\rm phot}$ correlation, which means
that galaxies with bluer $\beta$ have lower E$_{\rm s}$(B--V) for
$\beta_{\rm phot}$, and galaxies with redder $\beta$ have higher
E$_{\rm s}$(B--V) for $\beta_{\rm phot}$, compared to
$\beta_{\rm spec}$. Also, it should be noted that within the large
1$\sigma$ scatter in the E$_{\rm s}$(B--V) values, the difference
between these two $\beta$s is not large in all UV slope bins.  The
corresponding A$_{1600}$ (the dust absorption at 1600\AA) in
\figref{fig:beta_hist} (right panel) is estimated from
E$_{\rm s}$(B--V) using the Calzetti law
\citep[A$_{1600}$\,$\sim$\,10$\times$E$_{\rm s}$(B--V);][]{calz00}.
We have added best-fit relations from three different studies
  \citep{tali15,oteo14,buat12}. The \citet{oteo14} relation is for
  galaxies at $z$\,$\sim$\,2, while the \citet{buat12} relation is for
  galaxies at 1\,$\lesssim$\,$z$\,$\lesssim$\,2. The \citet{tali15}
  relation covers a larger redshift range, 1\,$<$\,$z$\,$<$\,3.  We do
  not see any trend (with redshift) between these relations and
  ours. Our relation is flatter compared to their relations and the
  difference is largest at redder UV slopes.  This difference between
  our relation and theirs could be due to the sample of SFGs used in
  these studies. All these studies use UV+FIR selected galaxies which
  are primarily dusty galaxies with FIR detection in 24$\mu$m and/or
  Herschel observations, while we do only UV selection and do not
  require FIR detection for our sample. This means that our sample
  does not include most of the dusty galaxies at these redshifts,
  which would make this relation much steeper with larger E(B--V)
  values and redder UV slopes. \citet{tali15} finds that the
  differences in various samples could also lead to a large dispersion
  ($\sim$0.7) in the dust-UV slope relation. Our future studies will
  specifically look at IR/FIR detections for these galaxies to
  investigate their dust properties.

In the remaining sections of this paper we use the spectroscopic UV
slope and refer to $\beta_{\rm spec}$ as $\beta$, unless stated
otherwise.  We are using $\beta_{\rm spec}$ in this paper because
  they have smaller measurement uncertainties, and the
  trends/relations explored in this paper are very similar for both
  spectroscopic and photometric UV slopes.

%
%

\subsection{$\beta$ versus M$_{\rm 1500}$}

\begin{figure*}
   \centering
   \includegraphics[scale=0.55]{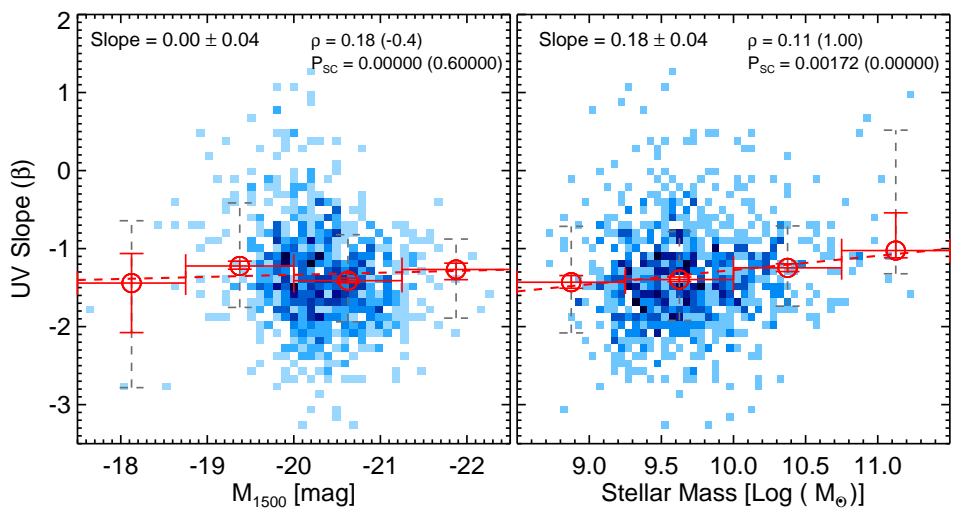}

   \caption{Spectroscopic UV slope versus the UV absolute magnitude
     (left) and the stellar mass (right). The density of points is
     color-coded as in previous figures, while the red open circles
     are the median-binned values for these parameters. The error bars
     in \emph{x} illustrate the sizes of the bins, while the errors in
     \emph{y} are $\pm$1$\sigma$ scatter (dashed error bars)
     corresponding to the range between the 16th and the 84th
     percentile values within each bin, while smaller solid error bars
     are the errors on the median values
     ($1.253\,\times\,\sigma/\sqrt{\rm N_{\rm gal}}$).  The dashed red
     lines show the best-fit linear relations. We find no correlation
     (slope\,=\,0.00$\pm$0.04) of $\beta$ with the UV absolute
     magnitude, while we see a positive correlation
     (slope\,=\,0.18$\pm$0.04) of $\beta$ with the stellar mass
     (redder UV slope for higher mass galaxies). The statistical
     significance from the Spearman correlation coefficient is shown
     to confirm these results.}

         \label{fig:mass_mag_beta}
   \end{figure*}

A color-magnitude relation has previously been observed at various
redshifts. The interpretation of this relation depends on which color
is used. If the color is between the rest-frame UV and the rest-frame
optical \citep[e.g.,][]{papo04} and spans the 4000\AA/Balmer break,
then it is strongly dependent on the stellar population age, while if
the color is blue-ward of the 4000\AA\ break \citep[e.g.,][]{fink12}
then it is more sensitive to the dust extinction than age. Here we use
rest-frame UV colors (i.e., UV spectral slope, $\beta$) to infer the
color-magnitude relation for SFGs at \ztwo.
\figref{fig:mass_mag_beta} shows the spectroscopic UV slope versus UV
absolute magnitude (M$_{\rm 1500}$). We estimate the slope,
$d\beta/dM_{\rm 1500}$, by fitting a linear model to the median-binned
values, $\beta=a+d\beta/dM_{\rm 1500} \times\ M_{\rm 1500}$. We find the
best-fitting solution to be $d\beta/dM_{\rm 1500}$\,=\,0.00$\pm$0.04.
Essentially no correlation of $\beta$ with absolute magnitude is
observed, which could imply that $\beta$ does not change with UV
luminosity for $\sim$L$^*$ galaxies.  The Spearman correlation
coefficient (\rsc) is 0.18 (-0.4) for all (binned) M$_{\rm 1500}$--$\beta$
values, which implies no correlation at a high significance
(P$_{\rm SC}$\,=\,0.60). The \rsc\ for all galaxies shows high
significance (P$_{\rm SC}$\,<\,10$^{-3}$) for a strong correlation but
it is misleading as it is strongly affected by a small number of
galaxies with significant redder UV slope.

Various studies have investigated the color-magnitude relation at high
redshifts ($z$\,$\gtrsim$\,2), and the results are
inconclusive. \citet{redd08} found weak-to-no correlation between dust
or UV color (related as shown in \figref{fig:beta_hist}) and $R$ mag
for UV-selected galaxies at 1.5\,$\lesssim$\,$z$\,$\lesssim$\,2.6. Similarly,
\citet{hein13} through their investigation of far-infrared/dust
properties of UV selected galaxies at $z$\,$\sim$\,1.5 found that the
average UV slope is mostly independent of the UV luminosity.  On the
other hand, \citet{bouw09} studied the relation between the UV slope and
$M_{UV}$ for a sample of $U$-band dropout galaxies at $z$\,$\simeq$\,2.5 and
found a positive color-magnitude
correlation. At higher redshifts ($z$\,$>$\,3), similar disagreements have
been observed for samples based on the Lyman break color selection, in
the sense that, some studies find no (or very weak) trends between
$\beta$ and UV magnitude \citep[e.g.,][]{ono10, fink12, dunl12,
  cast12}, while other studies find a strong color-magnitude relation
\citep[e.g.,][]{wilk11, bouw12,roge14,bouw14}.  \citet{bouw14} also
argue that this relation could be described by a double power-law with
a different slope for fainter magnitudes.  The differences in these
various studies could be due to many systematic and/or physical
reasons, such as selection of galaxy samples 
(especially at lower redshifts), UV magnitude and/or $\beta$
measurements, intrinsic scatter in $\beta$, dynamic range in UV
luminosity. Our study relies on a UV selected sample with
spectroscopic redshifts but it lacks a significant dynamic range in UV
luminosity as seen in \figref{fig:mass_mag_beta}. A detailed study of
a uniformly selected spectroscopic sample covering a large range in UV
luminosity is needed to understand the true nature of the
color-magnitude relation at high redshifts. At least for $\sim$L$^{*}$
($\sim$0.5L$^{*}$ -- 3.0L$^{*}$)
galaxies in a uniformly selected VUDS sample at \ztwo\ there appears to be
no correlation between $\beta$ and M$_{\rm 1500}$.
%
%

\subsection{$\beta$ versus Stellar Mass}

The relation between $\beta$ and M$_{\rm 1500}$ could be affected by the
biases and/or differences in the way different authors measure and
define M$_{\rm 1500}$ \citep[e.g.,][]{fink12}. On the other hand, there is
a well known correlation between M$_{\rm 1500}$ and the stellar mass such
that, stellar mass increases as brightness increases
\citep[e.g.,][]{star09,sawi12, hath13}, which we can exploit to better
understand the dependence of $\beta$ on M$_{\rm 1500}$. Therefore, we can
compare $\beta$ to the stellar mass as an alternate way to explore the
dependence of $\beta$ on M$_{\rm 1500}$.  \figref{fig:mass_mag_beta} shows
the correlation between the spectroscopic UV slope and the stellar
mass (M$_{*}$). We estimate the slope, $d\beta/dLog(M_{*})$, by
fitting a linear model to the median-binned values,
$\beta=a+d\beta/dLog(M_{*}) \times\ Log(M_{*})$.  We find the
best-fitting solution to be $d\beta/dLog(M_{*})$\,=\,0.18$\pm$0.04.  A
stronger correlation of $\beta$ with stellar mass is observed compared
to the $\beta$--M$_{\rm 1500}$ correlation, which suggests a redder UV
slope (more dust) for massive star-forming galaxies.  This is not
necessarily surprising as star-formation generally increases with
stellar mass in star-forming samples and in turn, the dust content
generally increases with increasing star-formation
\citep[e.g.,][]{lema14b}.  The Spearman correlation coefficient (\rsc)
is 0.11 (1.0) for all (binned) M$_{*}$-$\beta$ values, which implies
high significance (P$_{\rm SC}$\,$\sim$\,10$^{-3}$) for a strong positive
correlation. A similar trend between $\beta$ and stellar mass is also
observed at $z$\,$\gtrsim$\,4.0 \citep[e.g.,][]{fink12}.

The VUDS survey targets star-forming galaxies brighter than
$i_{AB}$\,$\sim$\,25~mag, so there are survey
limitation/incompleteness given the limited area of the survey, in the
sense that, we do not target many high mass galaxies
(Log(M$_*$)\,$\gtrsim$\,10.5), and we miss many fainter star-forming
galaxies with lower stellar masses (Log(M$_*$)\,$\lesssim$\,9.0)
because of the magnitude limit.  These survey limitations affect the
range in stellar mass range probed in our study, and hence, could
impact the exact nature of our $\beta$-stellar mass correlation. An
extensive study covering a large mass range is required to fully
understand this M$_{*}$--$\beta$ correlation.

The M$_{\rm 1500}$--$\beta$ and M$_{*}$--$\beta$ correlations are
  similar if we use $\beta_{\rm phot}$ instead of $\beta_{\rm spec}$.
  For $\beta_{\rm phot}$, the linear slope values for the
  M$_{\rm 1500}$--$\beta$ and M$_{*}$--$\beta$ relations are
  0.01$\pm$0.05 and 0.32$\pm$0.05, respectively.

%
\section{SFGs with and without \lya\ Emission}\label{lae}

The galaxies selected by their (strong) \lya-emission, are becoming an
important probe of galaxy formation, cosmic reionization, and
cosmology.  \lya\ emission is an important diagnostic of physical
processes in star forming galaxies, in particular at cosmological
distances, since \lya\ becomes the strongest emission line in the
UV-optical window at redshifts $z$\,$>$\,2.0.  These galaxies ---
because of their observed physical properties at higher redshifts ---
are thought to be the progenitors of present-day Milky Way type
galaxies \citep[e.g.,][]{gawi07}, though dissimilar results from
various \lya\ studies at different redshifts amplify the importance to
better understand these galaxies.  The SFG population in our sample is
divided into \lya\ emitters and non-emitters based on their rest-frame
\lya\ EW.

The \lya\ EWs for VUDS SFGs were measured as described in
\citet{cass15}. To summarize, we measured EWs of the \lya\ line
manually using the IRAF {\it splot} tool, while the uncertainties in
EW were estimated using the formalism from \citet{tres99}. The median
EW uncertainty varies as a function of the EW from $\pm$5\AA\ (for
weak absorbers and emitters) to $\pm$25\AA\ (for strong absorbers and
emitters). We first put each galaxy spectrum in its rest-frame based
on the spectroscopic redshift. Then, two continuum points bracketing
\lya\ are manually marked and the rest-frame EW is measured for
emission as well as the absorption lines. The line is not fitted with
a Gaussian, but the flux in the line is determined by simply summing
the pixels in the area covered by the line and the continuum. This
method allows the measurement of lines with asymmetric shapes (i.e.,
with deviations from Gaussian profiles), which is the shape observed
for most \lya\ lines.  The interactive method also allows us to
control the level of the continuum, taking into account defects that
may be present around the line. This approach produces more reliable
and accurate measurements compared to an automated `impartial'
measurement because the visual inspection is capable
of accounting for the complex nature of the \lya\ emission/absorption
line.

\begin{figure*}
   \centering
   \includegraphics[scale=0.525,angle=0]{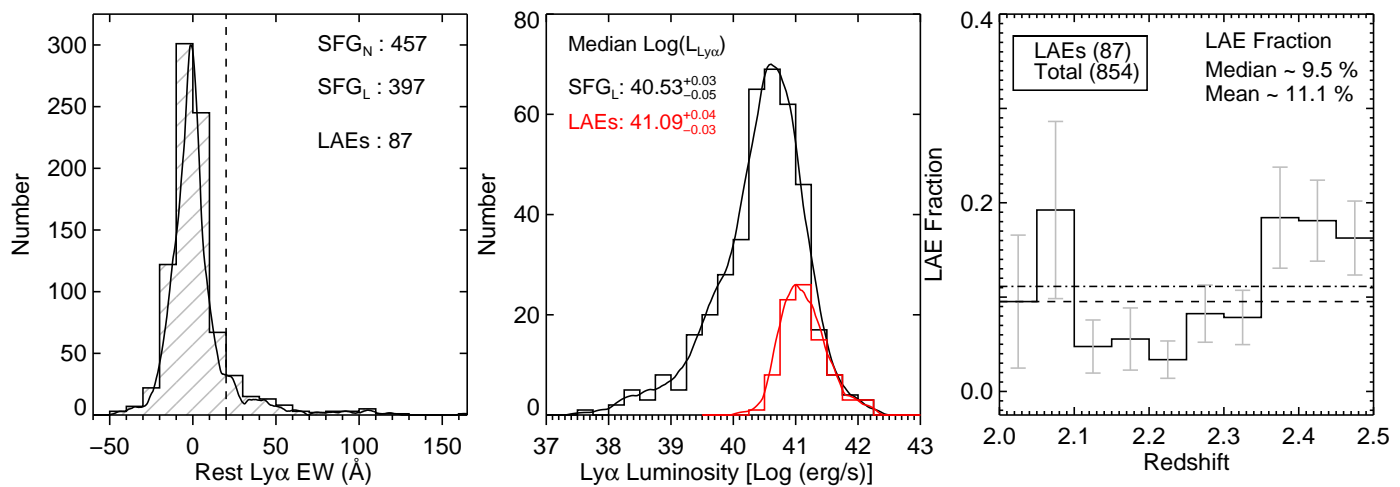}

  \caption{(Left) Rest-frame EW distribution for SFGs with \lya\ in
     emission (positive EWs) and \lya\ in absorption (negative EWs).
     The vertical dashed line indicates EW\,=\,20\AA, the adopted lower
     limit in \lya\ EW for LAEs (or strong \lya\ emitters). The solid
     curve is the KDE of the distribution. (Middle) Distribution of
     \lya\ luminosities for the \sfgl\ (black) and the LAE (red)
     samples. The KDE of the distribution is shown by a solid curve.
     (Right) Fraction of  LAEs  as a function
     of redshift for our sample. Approximately 10\% of SFGs at these
     redshifts are strong \lya\ emitters, consistent with previous
     \lya\ studies \citep[e.g.,][]{redd08,cass15} at similar redshifts.}

         \label{fig:ew_hist}
   \end{figure*}

\figref{fig:ew_hist} shows the rest-frame EW distribution for the
\lya\ line.  Here, positive EWs are emission lines, while negative EWs
are for absorption lines.  We divide the SFG population into three
sub-groups based on their \lya\ EW. The galaxies that show no \lya\ in
emission (EW\,$\le$\,0\AA) are defined as \sfgn, while the galaxies
with \lya\ in emission, irrespective of its strength (EW\,$>$\,0\AA),
are defined as \sfgl. The third group is for strong \lya\ emitters
(EW\,$\ge$\,20\AA) called LAEs.  The number of galaxies with \lya\
EW\,$\ge$\,20\AA\ is 87 and this number drops to 27 for galaxies with
EW\,$\ge$\,50\AA. Such a distribution is partially due to the fact
that our sample does not select most of the galaxies with very high EW
(EW\,$\gtrsim$\,50\AA), whose continuum is usually very faint
($\sim$27~mag) and which are usually detected in NB
imaging/emission-line selected surveys. This selection effect means
that we are probing \lya\ emitting galaxies with brighter continuum
compared to galaxies selected based on NB/emission-line surveys. We
also show the distribution of \lya\ luminosities, L(\lya), for \sfgl\
and LAEs in the middle panel of \figref{fig:ew_hist}. The median
L(\lya) of the \sfgl\ sample is $\sim$10$^{40}$ ergs/s, while the
median L(\lya) of the LAE sample is $\sim$10$^{41}$ ergs/s. The LAEs
discovered in the VUDS survey are significantly less luminous than the
\lya\ emitters found by most NB/emission-line surveys ($\sim$10$^{42}$
ergs/s) at similar redshifts \citep[e.g.,][]{guai10,hage14}.  The
fraction of LAEs as a function of redshift is shown in the right panel
of \figref{fig:ew_hist}. Approximately 10\% of SFGs at these redshifts
are strong \lya\ emitters, consistent with previous \lya\ studies at
$z$\,$\simeq$\,2 \citep[e.g.,][]{redd08}, and in agreement with the
general scenario that the number of LAEs increases as redshift
increases reaching $\sim$30-40\% at $z$\,$\simeq$\,6
\citep[e.g.,][]{star10,curt12,cass15}.
%
%
%

\begin{figure*}
   \centering
   \includegraphics[scale=0.525,angle=0]{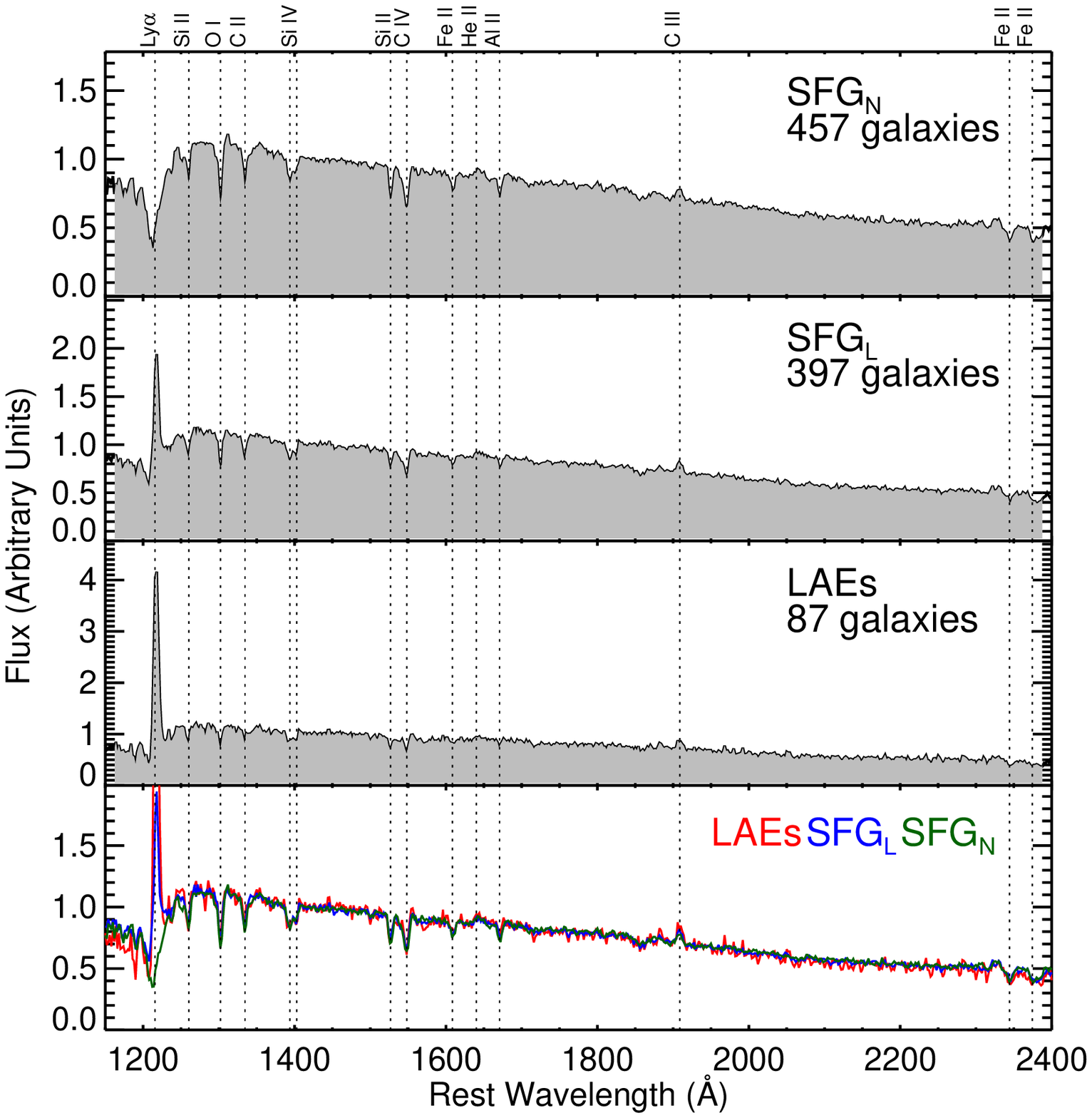}

   \caption{Stacked (median-combined) rest-frame UV spectra for \sfgn,
     \sfgl, and LAEs. The strong absorption and emission features are
     marked by the vertical dotted lines. The \sfgn\ spectrum shows
     strong absorption lines (Si, C, O, Fe, Al), but these absorption
     lines get weaker with increasing strength of \lya-emission. The
     bottom panel shows similarity in the spectroscopic UV slope for
     these samples.}

         \label{fig:stacks}
   \end{figure*}

   The composite spectra of the \sfgn, \sfgl, and LAE samples are
   shown in \figref{fig:stacks}. They are generated by stacking the
   rest-frame shifted individual spectra which are rebinned to
   dispersion of 2\AA\ per pixel (2\AA\,$\sim$\,5\AA/(1+$z$), where
   $\sim$5\AA\ is the native pixel scale of the LR Blue grism in the
   observed frame), and then the entire spectrum is normalized to the
   mean flux in the wavelength range of 1350–1450\AA\ before taking
   the median value of all galaxies in the composite at each
   wavelength. \figref{fig:stacks} shows the most prominent emission
   and absorption lines, whose detailed analysis will be conducted in
   a future paper. Here, it is worth pointing out that the median UV
   spectral slope (between 1400-2400\AA) does not show any appreciable
   difference among these three sub-samples  as shown in the bottom
   panel of \figref{fig:stacks}.

%

\begin{figure*}
   \centering
   \includegraphics[scale=0.65]{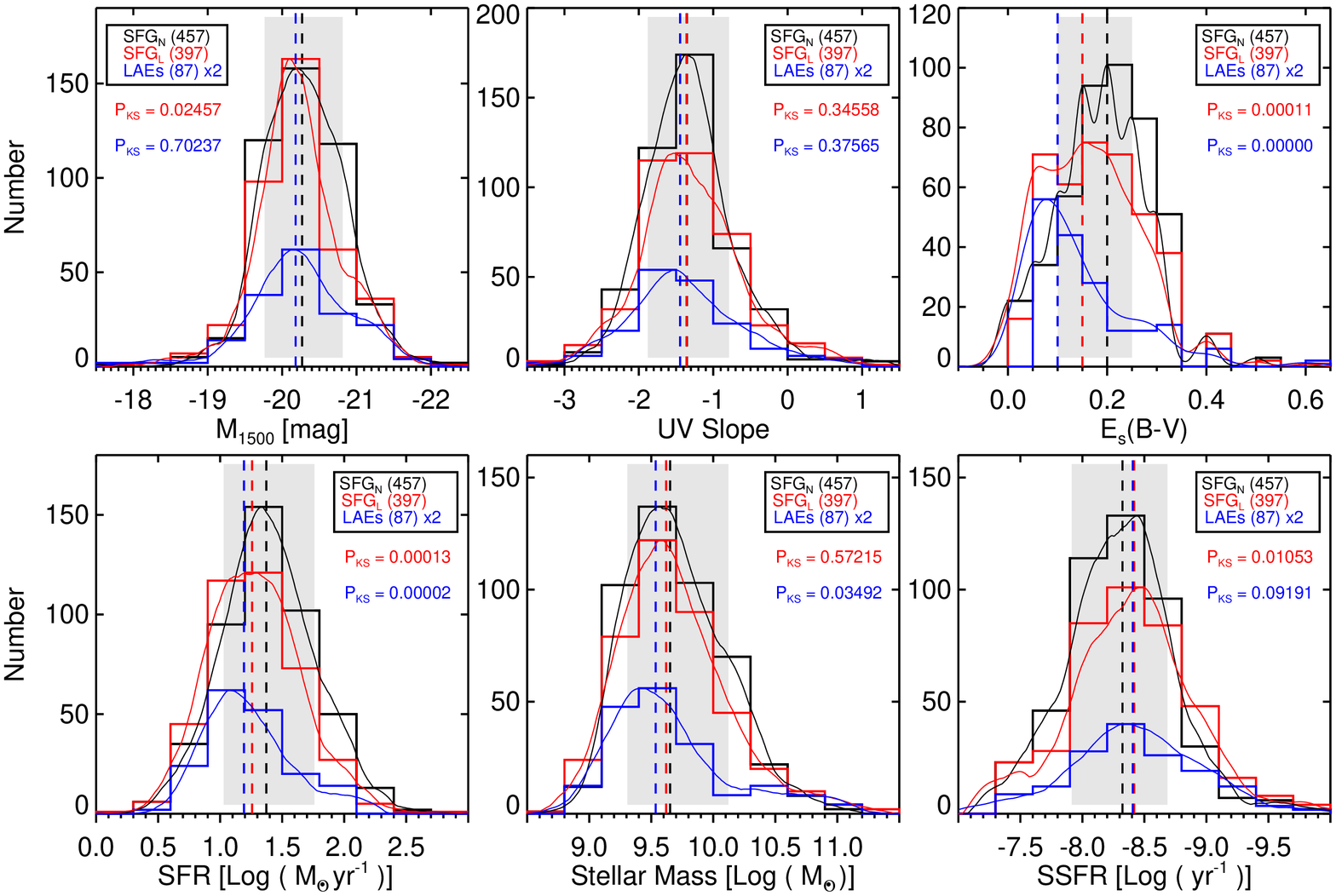}
 
   \caption{Comparison between \sfgn\ (black histograms), \sfgl\ (red
     histograms), and LAEs (blue histograms) as a function of stellar
     parameters. The total number of galaxies identified as \sfgn,
     \sfgl, and LAEs are shown in the legend. The median value of the
     distribution is indicated by a vertical dashed line. The solid
     curves are KDE of the distribution. The grey shaded regions cover
     $\pm$1$\sigma$ dispersion/scatter, which corresponds to the range
     between the 16th and the 84th percentile values, of the \sfgn\
     distribution.  The statistical significance based on the KS test
     (red for \sfgl, and blue for LAEs with respect to \sfgn) is shown
     to confirm any correlation. }
         \label{fig:sfg_compare}
   \end{figure*}

   The galaxies with strong \lya\ in emission, usually selected using
   NB imaging or serendipitous spectroscopy, show different physical
   properties compared to galaxies with no \lya\ in emission or LBGs
   \citep[e.g.,][]{gawi06,gawi07, lai08,cass11}. These two galaxy
   populations (LAEs and LBGs) are selected differently so the
   differences seen when comparing the two populations are mainly due
   to selection effects. Here, we use a UV-selected sample (selected
   based on UV continuum/magnitude) to investigate stellar population
   differences between \lya\ emitting galaxies and non-\lya\ emitting
   galaxies. A common selection for \lya\ emitters as well as
   non-emitters could reveal real intrinsic differences between these
   two classes of objects.  \figref{fig:sfg_compare} shows the
   comparison between \sfgn, \sfgl and LAEs. The total number of
   galaxies identified as \sfgl\ and \sfgn\ is large enough (457
   \sfgn, 397 \sfgl, 87 LAEs) to get robust statistics. On average,
   \sfgl\ (and LAEs) are less dusty, and have lower SFR compared to
   \sfgn. These differences are small, but they are significant
     because of the large sample used in this study. To understand
   the significance of these differences we use KS-statistics on these
   distributions. The P$_{\rm KS}$\,$\sim$\,10$^{-4}$ value for
   E$_{\rm s}$(B--V) and SFR distributions implies that the null
   hypothesis (similar distributions) is rejected at $>$99.9\% or
   $>$3$\sigma$ level. The differences in stellar mass, SSFR, M$_{\rm
     1500}$, and $\beta_{\rm spec}$ for \sfgn\ and \sfgl/LAEs are not
   statistically significant (i.e., the null hypothesis is rejected
   only at $\lesssim$2.5$\sigma$ level).  
    It is important to note that the KS-test does not consider
     errors/uncertainties in two quantities while comparing their
     distributions. We tested the robustness of these results by
     constructing 100 bootstrap samples by randomly extracting values
     from the confidence intervals of the best-fit stellar
     parameters. We re-ran the KS test on each of these 100 artificial
     samples with a goal to compare the stellar population parameters
     of the LAE and non-LAE populations. For E$_{\rm s}$(B--V) and SFR,
     a null hypothesis was consistently ruled out at $\sim$3$\sigma$
     level, while for other parameters, the null hypothesis rejection
     is at much lower significance ($\lesssim$2$\sigma$).  We also
     compare the median values of the best-fit stellar parameters for
     LAEs and non-LAEs, and we find similar significance for the
     median values of E$_{\rm s}$(B--V), SFR, and other parameters in
     these bootstrap samples.  These artificial samples show that even
     when accounting for the uncertainty in best-fit parameters, a
     significant difference exists for E$_{\rm s}$(B--V) and SFR, such
     that galaxies with \lya\ emission tend to be less dusty, and
     lower in SFR than galaxies with weak or no \lya\ emission.

%

   \tabref{table:1} summarizes average spectral and photometric
   properties of our \sfgn, \sfgl, and LAE samples.  Our results show
   that \sfgl\ (and LAEs) have --- on average --- lower dust content
   than \sfgn.  The differences in these two samples is statistically
   significant based on the K-S test results
   (P$_{\rm KS}$\,$\sim$\,10$^{-4}$), which says that distributions of
   E$_{\rm s}$(B--V) are different for \sfgn\ and \sfgl.  Though the
   difference in the distributions of these populations is
   statistically significant, the difference between their median
   values is small as shown in \tabref{table:1}. The median
   E$_{\rm s}$(B--V) values for \sfgl\ or LAEs and \sfgn\ differ only by
   $\sim$0.05 or $\sim$0.1 (i.e., by a factor of $\sim$1.3 or  2) which is
   smaller than or similar to the
   1$\sigma$ scatter in these distributions ($\sim$0.1).  This small
   difference in the dust content of \sfgn\ and \sfgl\ or LAE galaxies is
   consistent, within the large scatter, with the small difference in the
   median values of $\beta_{\rm spec}$, and a small difference in
   $\beta_{\rm phot}$ values ($\sim$0.1-0.2) between these 
   populations.  Such a small difference between the E$_{\rm s}$(B--V)
   values for \sfgl\ or LAEs and \sfgn\ is also consistent with the
   observations at higher redshifts. \citet{pent10} found small
   differences between E$_{\rm s}$(B--V) values for their samples of
   LBG$_N$ and LBG$_L$ at $z$\,$\sim$\,3.  A similar study at
   $z$\,$\sim$\,4 by \citet{pent07} agrees with this E$_{\rm s}$(B--V)
   trend for galaxies with and without \lya\ in emission.

%
\begin{table*}
\caption{Average properties of \sfgn, \sfgl, and LAEs derived from the SED-fitting}  
\label{table:1}      
\centering                                      
\begin{tabular}{c c c c}          
\hline\hline                        
Parameters$^a$ & \sfgn$^b$ & \sfgl$^c$ & LAEs$^d$ \\    
\hline                                   
    N$_{\rm gal}$ & 457 & 397 & 87 \\[2ex]      
    M$_{\rm 1500}$ (mag) & --20.27$^{+0.03\;(0.50)}_{-0.03\;(0.55)}$ & --20.18$^{+0.03\;(0.43)}_{-0.04\;(0.57)}$ & --20.18$^{+0.07\;(0.54)}_{-0.10\;(0.78)}$ \\[2ex]
    M$_{\rm 2300}$ (mag) & --20.56$^{+0.03\;(0.45)}_{-0.03\;(0.55)}$ & --20.45$^{+0.03\;(0.36)}_{-0.04\;(0.55)}$ &  --20.46$^{+0.05\;(0.38)}_{-0.09\;(0.66)}$ \\[2ex]
    \lya\ Luminosity (Log [erg/s]) & --- & 40.53$^{+0.03\;(0.52)}_{-0.05\;(0.75)}$ & 41.09$^{+0.06\;(0.41)}_{-0.04\;(0.31)}$ \\[2ex]
    UV slope (spectroscopic) & 	--1.35$^{+0.03\;(0.56)}_{-0.03\;(0.53)}$ & --1.36$^{+0.04\;(0.66)}_{-0.03\;(0.52)}$ & --1.45$^{+0.10\;(0.80)}_{-0.07\;(0.58)}$ \\[2ex]
    UV slope (photometric) & 	--1.30$^{+0.03\;(0.45)}_{-0.02\;(0.35)}$ & --1.35$^{+0.04\;(0.59)}_{-0.03\;(0.41)}$ & --1.58$^{+0.11\;(0.85)}_{-0.04\;(0.30)}$ \\[2ex]
    E$_{\rm s}$(B--V)  (mag) & 0.20$^{+0.00\;(0.05)}_{-0.01\;(0.10)}$ & 0.15$^{+0.01\;(0.10)}_{-0.01\;(0.10)}$ & 0.10$^{+0.02\;(0.15)}_{-0.01\;(0.05)}$  \\[2ex]
    SFR$_{\rm SED}$ (Log [$M_{\odot}\cdot yr^{-1}$]) & 1.37$^{+0.02\;(0.39)}_{-0.02\;(0.34)}$ & 1.26$^{+0.02\;(0.36)}_{-0.02\;(0.34)}$ & 1.19$^{+0.05\;(0.40)}_{-0.04\;(0.28)}$  \\[2ex]
    SFR$_{\rm UV}$ (Log [$M_{\odot}\cdot yr^{-1}$]) & 1.23$^{+0.03\;(0.56)}_{-0.03\;(0.57)}$ & 1.19$^{+0.04\;(0.59)}_{-0.04\;(0.60)}$ & 0.99$^{+0.12\;(0.91)}_{-0.09\;(0.65)}$  \\[2ex]
    Mass (Log [$M_{\odot}$]) & 9.65$^{+0.03\;(0.47)}_{-0.02\;(0.34)}$ & 9.62$^{+0.03\;(0.46)}_{-0.02\;(0.36)}$ & 9.54$^{+0.08\;(0.62)}_{-0.04\;(0.33)}$  \\[2ex]
    SSFR (Log [$yr^{-1}$]) & --8.32$^{+0.02\;(0.41)}_{-0.02\;(0.36)}$ & --8.42$^{+0.03\;(0.46)}_{-0.03\;(0.44)}$ & --8.41$^{+0.06\;(0.43)}_{-0.08\;(0.59)}$  \\[2ex]	
\hline   
                                          
\end{tabular}
\tablefoot{
\tablefoottext{a}{Median values of SED parameters and their quoted
  uncertainties ($1.253\,\times\,\sigma/\sqrt{\rm N_{\rm gal}}$) are shown. The
  $\pm$1$\sigma$ dispersion, corresponding to the range between the 16th and
     the 84th percentile values, are  also shown in the parenthesis}\\
\tablefoottext{b}{Galaxies with no Ly$\alpha$ in emission (EW\,$\le$\,0\AA)}\\
\tablefoottext{c}{Galaxies with Ly$\alpha$ in emission (EW\,$>$\,0\AA)}\\
\tablefoottext{d}{Galaxies with Ly$\alpha$ EW\,$\ge$\,20\AA}
}
\end{table*}

\figref{fig:sfg_compare} also shows that \sfgl\ (and LAEs) have --- on
average --- lower SED-based SFRs than \sfgn. The KS test results show
that distributions of the \sfgl\ and \sfgn\ samples are significantly
different i.e., P$_{\rm KS}$\,$\sim$\,10$^{-4}$, which is as
significant as the E$_{\rm s}$(B--V) comparison. The median SFR for the
\sfgl\ or LAE sample from the SED-fitting technique is
10$^{1.26}$\,$\sim$\,18 M$_{\odot}$ yr$^{-1}$ or
10$^{1.19}$\,$\sim$\,15 M$_{\odot}$ yr$^{-1}$, while the median SFR
for the \sfgn\ sample is 10$^{1.37}$\,$\sim$\,23 M$_{\odot}$
yr$^{-1}$. Therefore, the median SFR values for \sfgl\ or LAEs and
\sfgn\ differ only by a factor of $\sim$1.3 or 1.5, which is much
smaller in comparison to the 1$\sigma$ scatter of these distributions
($\sim$0.4 dex or a factor of $\sim$2).  To assess the effect of
SED-fitting parameters on the SFRs \citep[e.g.,][]{kusa15}, we also
compute SFR using the UV luminosity (M$_{\rm 1500}$) corrected for the
dust using the $\beta$--A$_{1600}$ relation from \citet{meur99}. The
median values for the UV-based SFRs (SFR$_{\rm UV}$) are shown in
\tabref{table:1} and they are very similar to the SED-based SFRs. The
median SFR$_{\rm UV}$ for the \sfgl\ or LAE sample is
10$^{1.19}$\,$\sim$\,15 M$_{\odot}$ yr$^{-1}$ or
10$^{0.99}$\,$\sim$\,10 M$_{\odot}$ yr$^{-1}$, and for the \sfgn\
sample is 10$^{1.23}$\,$\sim$\,17 M$_{\odot}$ yr$^{-1}$. Therefore,
the difference between SFR$_{\rm UV}$ values of these two populations
is very small and agrees very well with the difference quoted for the
SED-based SFRs.  This is consistent with the fact that the UV absolute
magnitudes (M$_{\rm 1500}$) of galaxies with and without \lya\ in
emission are very similar as shown in the \tabref{table:1}.  This
result is consistent with similar studies at $z$\,$\sim$\,3--4
\citep[e.g.,][]{pent07,pent10}.
  
\figref{fig:sfg_compare} does not show any significant difference
between the median values of stellar mass ($\sim$0.1 dex or a factor
of $\sim$1.2). This result is in contrast to higher redshift studies
at $z$\,$\sim$\,3--4 \citep[e.g.,][]{pent07,pent10}, where they find
that galaxies without \lya\ in emission are more massive (by a factor
of $\sim$2--5) compared to galaxies with \lya\ in emission. The
difference between these studies could be due to the sample selection,
in that the stellar mass probed in those studies by the galaxies at
$z$\,$\sim$\,3--4 are much lower (down to $\sim$10$^8$ M$_{\odot}$)
compared to the VUDS sample (down to $\sim$10$^9$ M$_{\odot}$). The
lowest mass regime at $z$\,$\sim$\,3--4 is mostly populated by
galaxies with \lya\ in emission which results in this large difference
in stellar masses for \lya\ emitters and non-emitters at these
redshifts. We do not probe this low mass regime between 10$^8$ and
10$^9$ M$_{\odot}$ at \ztwo\ because of our continuum magnitude limit
of $i_{AB}$\,$\lesssim$\,25~mag. Therefore, we conclude that, within
the luminosities probed by VUDS, we do not see any significant
difference between the stellar mass of galaxies with and without \lya\
in emission. We will investigate any redshift evolution in these
correlation in our future VUDS studies, which will extend these
measurements to higher redshifts ($z$\,$\gtrsim$\,3).

\section{\lya\ EW and Stellar Population Properties}\label{lae_sp}

\begin{figure*}
   \centering
   \includegraphics[scale=0.65]{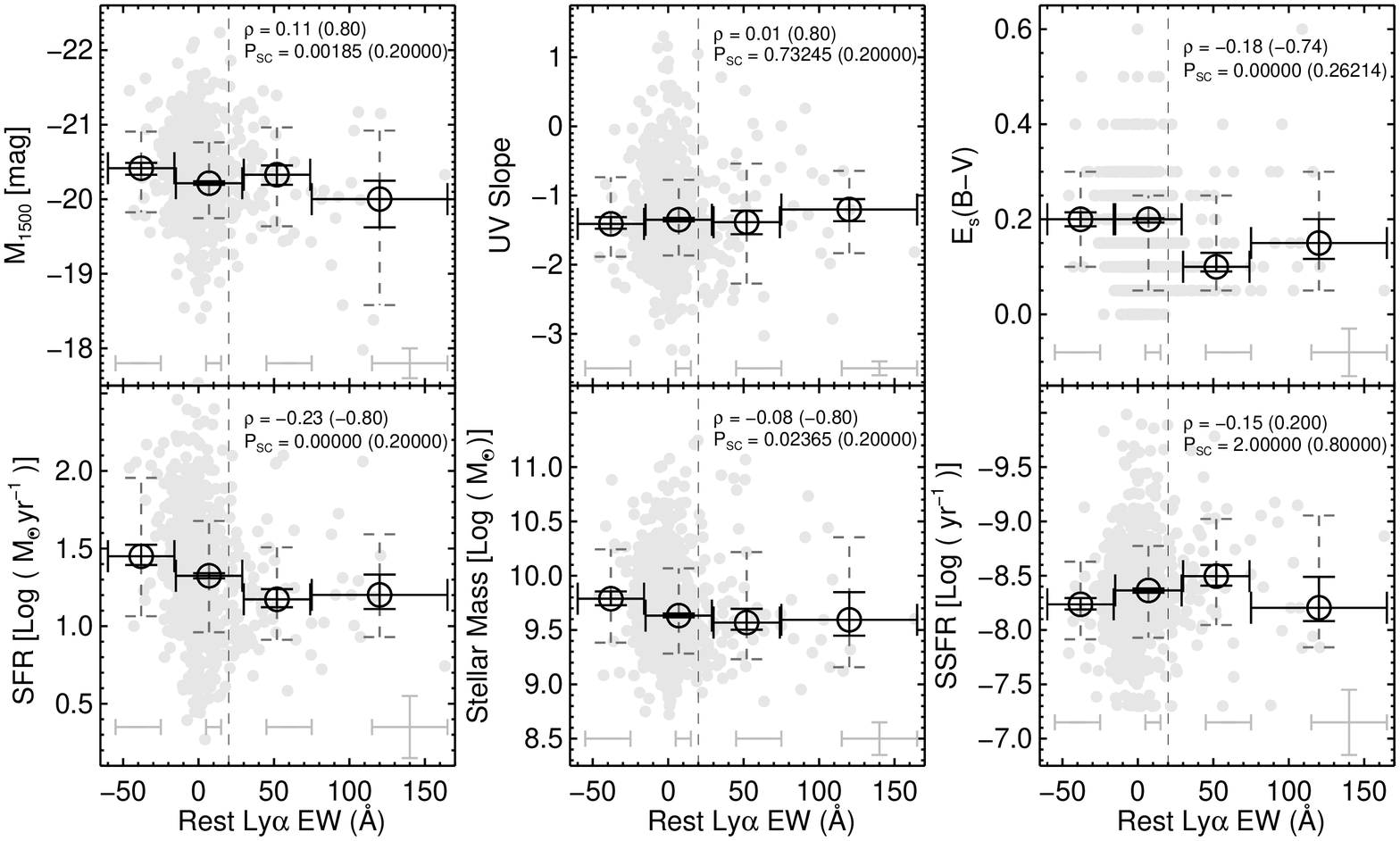}

   \caption{Correlation between rest-frame \lya\ EW and best-fit
     stellar population parameters.  The vertical dashed line
     indicates EW\,=\,20\AA, the adopted lower limit in \lya\
     equivalent width for LAEs.  The size of the last bin is increased
     with respect to that of the other bins to increase the number of
     galaxies in that bin. The light grey points are individual
     measurements while the black circles are binned medians. The
     error bars in \emph{x} illustrate the sizes of the bins, while
     the errors in \emph{y} are $\pm$1$\sigma$ scatter (dashed error
     bars) corresponding to the range between 16th and the 84th
     percentile values within each bin, while smaller solid error bars
     are the errors on the median values
     ($1.253\,\times\,\sigma/\sqrt{\rm N_{\rm gal}}$).  The
     statistical significance from the Spearman correlation
     coefficient for all galaxies (median-binned values) is shown to
     confirm these correlations. Average uncertainties in these
     measurements are plotted in the bottom-right corner. The median
     uncertainties in the EW measurement varies as a function of EW as
     shown by different \emph{x} error bars.}

         \label{fig:ew_compare}
   \end{figure*}

To further investigate properties of \sfgn\ and \sfgl, we explore the
correlation between the rest-frame \lya\ EW and stellar population
parameters, such as M$_{\rm 1500}$, UV slope, E$_{\rm s}$(B--V), SFR,
stellar mass, and SSFR.  \figref{fig:ew_compare} shows the correlation
between \lya\ EW and best-fit SED/spectral parameters.  It should be
noted that the Spearman correlation is measured from all galaxies as
well as from the median-binned values.  The E$_{\rm s}$(B--V) and SFR
show moderate-to-weak monotonic correlation (\rsc\,$\sim$\,--0.2 and
P$_{\rm SC}$\,<\,10$^{-4}$) with \lya\ EW.  The significance of \rsc\
as measured by the number of standard deviations by which the
sum-squared difference of ranks deviates from its null-hypothesis
expected value is $\sim$5$\sigma$ for all E$_{\rm s}$(B--V) values and
$\sim$6$\sigma$ for all SFR values.  The stellar mass, M$_{\rm 1500}$,
and $\beta$ show much weaker correlation with EW
(\rsc\,$\lesssim$\,0.1). The significance of \rsc\ as defined above is
$\lesssim$3$\sigma$ for all values of these parameters.  The trends
observed between the best-fit stellar parameters and EWs are in
general agreement with the median values of the stellar parameters
obtained from the distributions of \sfgn\ and \sfgl\ in
\figref{fig:sfg_compare}.
%

   With increasing rest-frame \lya\ EW, we see a weak but
     significant trend in that, galaxies are less dusty, and less
     star-forming, as indicated by a decrease in SFR by about
     $\sim$0.2 dex from non-LAEs to LAEs. While the average EW varies
   from --38.0\AA\ to 120\AA, the median E$_{\rm s}$(B--V) decreases
   from 0.20 to 0.15, the median SFR$_{SED}$ decreases from 28
   M$_{\odot}$ yr$^{-1}$ to 16 M$_{\odot}$ yr$^{-1}$, the median
   stellar mass decreases from 6.2$\times$10$^9$ M$_{\odot}$ to
   3.9$\times$10$^9$ M$_{\odot}$, and the median SSFR decreases from
   5.9$\times$10$^{-9}$ yr$^{-1}$ to 6.3$\times$10$^{-9}$
   yr$^{-1}$. All these differences in stellar parameters as a
   function of \lya\ EW are small compared to the scatter observed in
   each EW bin, as measured by 1$\sigma$ dispersion (see
   \figref{fig:ew_compare}).

   These results demonstrate that there is a statistically significant
   correlation between \lya\ EW and stellar parameters, such as SFR
   and E$_{\rm s}$(B--V), while no significant correlations are found
   for stellar mass, M$_{\rm 1500}$, and $\beta$. The differences
     we observe are small compared to the large scatter in their
     distributions.  This outcome is true whether or not we focus on
   all galaxies with \lya\ in emission or only strong \lya\ emitters
   (LAEs). This is in contrast with results from NB or emission-line
   selected LAE studies \citep[e.g.,][]{gawi06,gawi07,lai08,
     guai11,hage14}.  These studies find lower dust, lower stellar
   mass and higher SSFR for galaxies with strong \lya\ in
   emission. The LAE samples used in these emission-line/NB-selected
   surveys have \lya\ luminosities in the range of $\sim$10$^{42}$
   ergs/s, which is an order of magnitude more luminous than VUDS
   LAEs, as shown in \figref{fig:ew_hist}. It is also worth
   emphasizing that VUDS is a UV continuum selected survey and
   galaxies are targeted based on their photometric redshifts and not
   on the strength of the \lya\ line.  We have 27 galaxies beyond
   rest-frame \lya\ EW\,=\,50\AA, so by selection, our sample does not
   include extremely low mass ($\sim$10$^8$ M$_{\odot}$) galaxies
   usually selected from emission-line/NB techniques.  Moreover, among
   the emission-line/NB-selected LAEs, those that have lower \lya\
   luminosities, have similar SFR as normal star-forming galaxies but
   lie on the lower mass end of the star-forming main sequence
   \citep[e.g.,][]{kusa15}. Hence, the difference in the sample
   selection could play a significant role in such a comparison as we
   could be comparing two intrinsically different populations of
   galaxies.

     The comparison of stellar population properties of galaxies with
     and without \lya\ emission has also been done on UV-selected
     galaxies at $z$\,$\simeq$\,2 and higher, and our results are
     broadly consistent with the literature. \citet{pent10} found
     weak/no correlation between SFRs of galaxies with and without
     \lya\ emission at $z$\,$\simeq$\,3 but we find a correlation
     (with a small difference in SFR) possibly because of the large
     statistics in the VUDS sample. With a smaller number of galaxies
     (130) in the Pentericci et al. sample such a detection was
     difficult to observe.  Our finding of a very small decrease in
     the SED-based dust content (i.e., E$_{\rm s}$(B--V)) of these two
     populations agrees well with the \citet{pent10} study. At
     $z$\,$\simeq$\,3, \citet{korn10} find that LAEs have lower
     E$_{\rm s}$(B--V) and SFRs compared to non-LAEs. The difference in
     these two stellar population parameters for LAEs and non-LAEs, as
     found by Kornei et al., is much larger than what we observe, but
     show similar decreasing trends in these two quantities. This
     trend of lower dust and SFR for strong \lya\ emitters has also
     been observed at $z$\,$\simeq$\,2 by \citet{redd06}. The study of
     \citet{korn10} agrees well with our result that there is no
     significant difference between stellar masses of LAEs and
     non-LAEs. Therefore, based on the comparison between \sfgl/LAEs
     and \sfgn, as well as \lya\ EW correlations with stellar
     parameters, we suggest that our LAE sample shows a small decrease
     in E$_{\rm s}$(B--V) and SFR but otherwise their stellar
     populations are not very different from the galaxies that do not
     show \lya\ in emission.

\begin{figure*}
   \centering
   \includegraphics[scale=0.65]{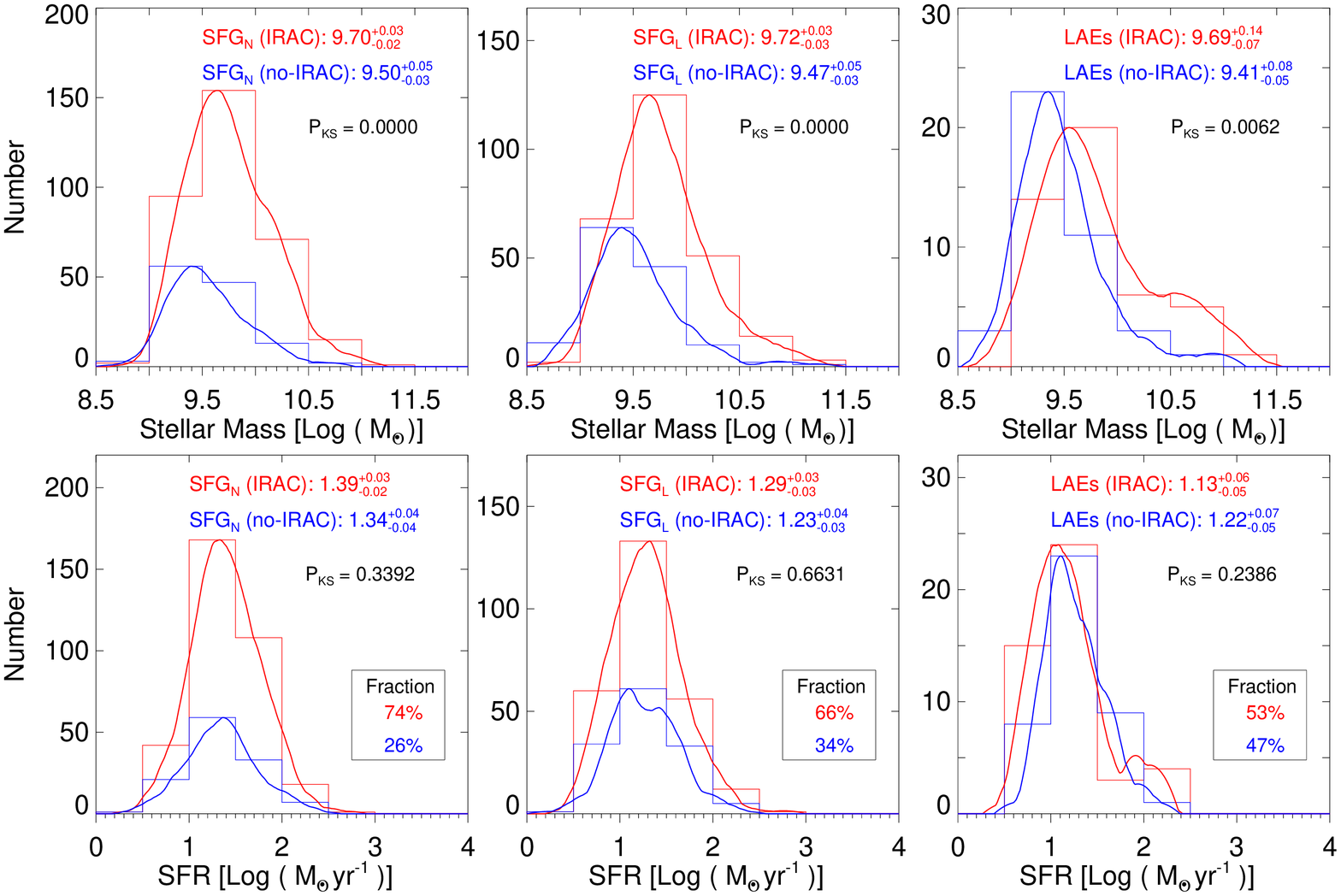}

   \caption{Comparison between Spitzer/IRAC 3.6$\mu$m-detected (red;
     brighter than m$_{3.6}$\,$\sim$\,25 mag) and IRAC-undetected
     (blue; fainter than m$_{3.6}$\,$\sim$\,25 mag) galaxies as
     function of stellar mass and SFR for \sfgn, \sfgl, and LAEs. The
     fraction of galaxies detected/un-detected in IRAC are shown in
     the legend. The solid curve is the KDE of the distribution.  The
     statistical significance based on the KS test is shown to confirm
     the observed correlation. }
         \label{fig:sfg_3.6m}
   \end{figure*}

     \citet{lai08} studied NB-selected LAEs at $z$\,$\simeq$\,3 by
     dividing them into Spitzer/IRAC-detected (brighter than
     m$_{3.6}$\,=\,25.2 mag) and IRAC-undetected (fainter than
     m$_{3.6}$\,=\,25.2 AB mag) samples. They found that
     IRAC-undetected LAEs were less massive and younger compared to
     IRAC-detected LAEs, concluding that IRAC-detected LAEs are a more
     evolved population similar to brighter/massive UV-selected LAEs
     \citep[e.g.,][]{shap01,shap03}. To investigate this trend for
     VUDS LAEs, we examined our LAEs as a function of magnitude in the
     Spitzer/IRAC 3.6$\mu$m band.  We find that 74\% of \sfgn, 66\% of
     \sfgl, and 53\% of LAEs have detections in the Spitzer/IRAC
     3.6$\mu$m band down to $\sim$25 mag. The LAE fraction with IRAC
     detection in our sample is much larger than the $\sim$30\% found
     in \citet{lai08}, which is a direct consequence of the difference
     in the sample selection between the two studies.  The
       NB-selected LAEs typically have very faint continuum in all
       optical and NIR bands as they are mostly lower mass/younger
       galaxies compared to normal star-forming
       galaxies. IRAC-detection is only possible in NB-LAEs if they
       host more evolved population which makes continuum 
       brighter in the red bands. We do not expect many NB-selected
       LAEs to be more evolved/massive and brighter in IRAC bands,
       therefore their IRAC-detected fractions are more likely to be
       small  compared to other star-forming galaxies.  A
     similar study at $z$\,$\simeq$\,2 \citep{guai11}, also found that
     $\sim$25\% of NB-selected LAEs had IRAC detection down to
     m$_{3.6}$\,$\sim$\,24.5 AB mag. We find a higher detection rate
     (51\%) of IRAC-detected LAEs in our sample using the same
     limiting magnitude of 24.5 mag as Guaita et al.
     \figref{fig:sfg_3.6m} shows the comparison of IRAC-detected and
     IRAC-undetected galaxies in all three samples as a function of
     stellar mass, and SFR. The stellar mass correlations show that,
     in all three samples, IRAC-detected galaxies are more massive
     than IRAC-undetected galaxies. These correlations are
     statistically significant based on their P$_{\rm KS}$ values
     $<$\,10$^{-2}$.  We do not see any significant correlation for
     SFR in all three samples, both in terms of P$_{\rm KS}$ and their
     median values.

   These comparisons show that the VUDS LAE sample has a larger
   fraction of galaxies with IRAC detection compared to NB-selected
   LAEs at $z$\,$\simeq$\,2--3, and that IRAC-detected LAEs are more
   likely to host evolved populations compared to IRAC-undetected
   LAEs.  A large fraction of UV-selected LAEs have similar
     masses as \sfgn\ galaxies but with lower SFRs suggesting that
     these LAEs host more evolved populations.  The low fraction of
   IRAC detection in NB-selected LAEs implies that these galaxies are
   less evolved compared to UV-selected LAEs.

%

\section{Results and Discussion}\label{results}

We use a large spectroscopic sample of 854 SFGs at \ztwo\ from VUDS to
investigate their spectral and photometric properties. The VUDS
spectra were used to measure the UV spectral slope and \lya\ EW. The
SED fitting on multi-wavelength photometric data using \texttt{Le
  PHARE} provided best-fitting stellar parameters, such as stellar
mass, M$_{\rm 1500}$, E$_{\rm s}$(B--V), SFR, and SSFR. The VUDS
observations extend to a fairly large area of $\sim$1~deg$^2$, which
implies that our sample spans a large range in SFR ($\sim$3-150
M$_{\odot}$ yr$^{-1}$) and stellar mass ($\sim$5$\times$10$^8$ to
10$^{11}$ M$_{\odot}$). The VUDS \ztwo\ sample covers a substantial
range in UV absolute magnitude around M$^*$ ($\pm$1~mag), the
characteristic magnitude at these redshifts.  We divide SFGs into
three sub-groups, \sfgn, \sfgl\ and LAEs, based on their rest-frame
\lya\ EW. We find that the fraction of LAEs is $\sim$10\% at these
redshifts, which is consistent with previous studies
\citep[e.g.,][]{redd08}, and is in accordance with the general picture
where the \lya\ fraction increases with redshift
\citep[e.g.,][]{star10,curt12,cass15}. The \lya\ fraction apparently
starts to decrease at higher redshift \citep[$z$\,$\gtrsim$\,6.5;
e.g.,][]{pent11}, where large amounts of neutral hydrogen start to
affect the visibility of \lya\ line indicating the onset of the
reionization epoch.

The UV spectral slope $\beta$ can be used to derive the dust
attenuation for local starburst galaxies
\citep[e.g.,][]{calz94,leit99} and has been adopted as a dust
indicator for high redshifts galaxies
\citep[e.g.,][]{noll04,hath08b,fink12,bouw14}. The main reason is that
the intrinsic UV slope depends only weakly on metallicity and stellar
populations for star-forming galaxies \citep[e.g.,][]{heck98,leit99}.
In addition, $\beta$ shows a strong correlation with the
L$_{\rm IR}$-to-L$_{\rm UV}$ ratio, a standard dust indicator
\citep[e.g.,][]{meur99,redd10}. On average, the spectroscopic UV slope
for SFGs at \ztwo, measured using VUDS spectra,  is comparable to
  the photometric $\beta$ measured using multiple photometric bands,
  and have smaller measurement uncertainties. The measured UV slope
--- spectroscopic and photometric --- is in accordance with the
evolutionary trend of $\beta$ with redshift, which shows that lower
redshift galaxies have, on average, redder UV slope compared to higher
redshift galaxies \citep[e.g.,][]{bouw12,hath13}.  With dust being the
major factor influencing the UV slope, this implies that lower
redshift galaxies have more dust compared to higher redshift
galaxies. We use the spectroscopic $\beta$ measurements to explore its
correlation with M$_{\rm 1500}$ and stellar mass. Comparing $\beta$ to
the stellar mass we find a significant correlation, in the sense that
massive galaxies are redder, which is in general agreement with the
higher redshift measurements of \citet{fink12}. We find no correlation
between M$_{\rm 1500}$ and $\beta$. This result is at variance with
what found by some authors at higher redshifts
\citep[e.g.,][]{bouw12,roge14}, but it is similar to other studies
which do not find any significant correlation between $\beta$ and
M$_{\rm 1500}$ \citep[e.g.,][]{fink12, cast12,hein13}. It is vital to
note that this correlation could be affected by the biases and/or
differences in the way different authors measure M$_{\rm 1500}$ and/or
$\beta$ but most importantly the dynamic range in M$_{\rm 1500}$ plays
a crucial role. Extensive studies exploring a larger range in
M$_{\rm 1500}$ and stellar masses and investigating different biases
in these correlations will shed more light on the true nature of these
relations and provide a better physical understanding.

There are several studies investigating the stellar population of
UV-selected LAEs at $z$\,$\simeq$\,2--3
\citep[e.g.,][]{shap01,shap03,erb06,redd08,korn10}.   We find that
  LAEs in our sample have lower dust content, lower SFR and lower SSFR
  compared to non-emitters. These differences are very small compared
  to the large scatter in these SED-based parameters but they are
  significant because of the large sample size.  \citet{shap01} and
\citet{korn10} found that LAEs at $z$\,$\simeq$\,3 have lower SFRs,
lower dust and lower SSFR compared to LBGs, which is consistent with
our observed trends, but we find much smaller differences in these
stellar parameters for our galaxies.  We find only weak or no
correlation of stellar mass with \lya\ EW for a sample with median
Log(M$_*$)\,=\,9.64 and median Log(SFR)\,=\,1.32.  \citet{korn10}
studied a large ($\sim$300) sample of UV-selected galaxies,  with
  and without \lya\ emission, with a median stellar mass of
Log(M$_*$)\,=\,9.92 and a median SFR of Log(SFR)\,=\,1.57, both
slightly higher than for our sample. They found that there is no
significant difference in the stellar mass between these two
populations in accordance with our weak-to-no correlation of stellar
mass with the \lya\ EW.  The results obtained here show no
\emph{strong} correlation or large differences between stellar
parameters as a function of the \lya\ EW,
which is consistent with the small sample study of \citet{redd08}
using UV-selected galaxies at $z$\,$\simeq$\,2--3.  It is clear that
the properties of galaxies with and without Ly$\alpha$ in emission are
pretty similar suggesting that the two populations, at least for
typical ($\sim$L$^{*}$) galaxies and to the level of detail we are
able to probe, are roughly comprised of similar galaxies.
\citet{erb06} used composite spectra of UV-selected LAEs and non-LAEs
at $z$\,$\simeq$\,2 and found that galaxies with strong \lya\ EW had
lower stellar masses. Our results on SFR and dust are broadly
consistent with previous studies, but the lack of a strong correlation
between stellar mass and EW needs to be explored with a sample that
spans a larger dynamic range in stellar mass as lower mass galaxies
could strongly impact the stellar mass correlation between
\lya\ emitters and non-emitters.

In comparison to UV-selected LAEs, LAEs selected based on the
emission-line/NB technique at $z$\,$\simeq$\,2 and beyond have very
low masses $\sim$10$^8$ M$_{\odot}$ compared to $\sim$10$^{10}$
M$_{\odot}$ for non-LAEs \citep[e.g.,][]{gawi07, fink07,lai08,
  guai11,varg14}. Also, these LAEs are much more luminous in \lya\
($\sim$10$^{42}$ ergs/s) than VUDS LAEs, which are on average an order
of magnitude less luminous ($\sim$10$^{41}$ ergs/s). We do not find
such low mass, high \lya\ luminosity LAEs in our sample mainly because
these strong emitters have low continuum luminosities and are not
selected by the VUDS magnitude selection. In fact, such galaxies would
be missed in almost all UV continuum selected samples
\citep[e.g.,][]{shap01, redd08,korn10}, as emission-line/NB LAEs probe
a different luminosity range compared to magnitude-limited
samples. However, \citet{korn10} have argued that when NB-selected
LAEs are restricted to luminosities probed by continuum selected
samples, both samples are statistically very similar, which is
consistent with the result put forward by the study of \citet{verh08},
where bright NB-selected LAEs represent the same population as
continuum selected LAEs. Therefore, UV-selected LAEs and NB-selected
LAEs need to probe similar luminosities for proper comparison between
these two populations.

In the literature there are two possible scenarios under which
galaxies emit \lya\ during the evolution process. Studies of faint
NB-selected LAEs \citep[e.g.,][]{gawi06,gawi07,lai08} have argued that
LAEs represent the beginning of an evolutionary sequence of galaxy
formation. According to these studies, the main reason behind this
argument is that LAEs have lower dust content, lower stellar mass,
younger stellar ages, and higher SSFRs compared to non-LAEs which
means that LAEs are building up their stellar mass at a faster rate
than non-LAEs through mergers and/or star-formation episodes. On the
other hand, studies of brighter continuum selected LAEs
\citep[e.g.,][]{shap01,shap03,korn10} have suggested that LAEs
represent a later stage in the evolutionary sequence. These studies
argue that LAEs have lower dust content, lower SFR, older stellar
ages, and lower SSFRs compared to non-LAEs which means that LAEs are
more quiescent than non-LAEs. This could be the result of strong
outflows from supernovae and massive star winds, which expel both gas
and dust from a young, dusty non-LAE. These arguments raise a
question: are these two scenarios two different phases of the
evolution process or do these two possibilities describe a single
phase?  \cite{lai08} investigated the stellar populations of 162
NB-selected LAEs at $z$\,$\simeq$\,3.1 by dividing them in two groups
based on their Spitzer/IRAC flux.  They find that $\sim$70\% of the
LAEs are undetected in 3.6$\mu$m down to 25.2 mag. Based on their
stacking analysis, they find a clear difference between these two
divided samples. The average stellar population of the IRAC-undetected
sample had an age of $\sim$200 Myr and a mass of $\sim$3$\times$10$^8$
M$_{\odot}$, consistent with the scenario that LAEs are mostly young
and low-mass galaxies. On the other hand, the IRAC-detected LAEs were
on average significantly older and more massive, with an average age
of $\sim$1 Gyr and mass of $\sim$10$^{10}$ M$_{\odot}$. The stellar
populations of the IRAC-detected sample of \citet{lai08} are very
similar to those of continuum selected brighter LAEs
\citep[e.g.,][]{shap01,shap03,korn10}.  Similar results are also found
in the study of \citet{guai11} at $z$\,$\simeq$\,2.1, in that, the
fraction of IRAC-detected LAEs is $\sim$25\% down to 24.5 mag and
these galaxies are more massive/evolved than those that are faint in
IRAC.  We find larger fraction of IRAC-detected LAEs ($\sim$50\%)
compared to NB-selected LAEs ($\sim$30\%) at $z$\,$\simeq$\,2--3,
which could imply that UV-selected LAEs are more evolved compared to
NB/emission-line selected LAEs. Also, the IRAC-detected LAEs have
higher stellar mass, similar to IRAC-detected non-LAEs, but with lower
SFR which is consistent with the more evolved nature of these
galaxies.  The scenario in which LAEs are more evolved than non-LAEs
is plausible if strong outflows in non-LAEs destroy or remove dust and
gas from the galaxies, allowing \lya\ photons to escape.  Our results
from direct comparison of galaxies with and without \lya\ emission
show small differences in stellar populations, such that galaxies with
\lya\ emission have lower SFR and dust content than non-\lya\ emitting
galaxies. This result broadly agrees with previous studies involving
UV-selected galaxies at $z$\,$\simeq$\,3
\citep[e.g.,][]{shap03,korn10}, though we observe small difference
between physical parameters of \lya\ emitters and non-emitters.
Past studies of LAEs \citep[e.g.,][]{verh08, nils11, acqu12},
suggest that LAEs have a large range in stellar population properties
implying that a strong LAE phase either is a long duration phase
\citep[$\sim$1 Gyr;][]{lai08} or is recurring in SFGs.

Studies focusing on comparing stellar populations of galaxies with and
without \lya\ emission have shown that various effects, including the
sample selection, continuum and \lya\ luminosities, measurements of
stellar parameters, and SED fitting assumptions play an important
role in how we compare \lya\ emitters and non-emitters. It is also
essential to understand that these comparisons and correlations are
not universal and could change with redshift.  Extending these studies
to higher redshifts, where the fraction of \lya\ emitters increases,
is essential to generate a self-consistent picture of how LAEs and
non-LAEs form and evolve.  In future studies, we will extend such an
analysis to $z$\,$\simeq$\,3--6 using the VUDS data, to better
understand how these LAE properties/trends evolve with redshift
from $z$\,$\sim$\,6 to $z$\,$\sim$\,2.


\section{Summary}\label{summary}

In this paper, we have investigated the spectro-photometric properties of a
large sample of SFGs at \ztwo\ that were selected from the VIMOS Ultra-Deep
Survey in the ECDFS, COSMOS, and VVDS fields. These galaxies were
targeted because of their photometric redshifts, and are therefore UV
continuum-selected galaxies. The VUDS spectra were used to measure the
UV spectral slope ($\beta$) and \lya\ EWs, while we used deep
multi-wavelength observations in these extensively observed fields to
derive physical parameters (stellar mass, SFR, E$_{\rm s}$(B--V),
M$_{\rm 1500}$, SSFR) from the SED fitting process. We compared
\lya\ emitters and non-emitters using these parameters and also
explored correlations of these parameters with \lya\ EW. Our results
can be summarized as follows:

$\bullet$  We obtain reliable measurements of  spectroscopic UV
  slopes, which are --- on average --- similar to the photometric
  measurements of $\beta$, and have smaller measurement
  uncertainties.  The median values of $\beta_{\rm spec}$ and
$\beta_{\rm phot}$ are consistent with each other and with the general
picture in which UV slopes get redder with decreasing redshift,
implying higher dust content at lower redshifts. We find a significant
correlation between $\beta$ and SED-based dust indicator
E$_{\rm s}$(B--V). We observe no correlation between $\beta$ and
M$_{\rm 1500}$, while a strong correlation between $\beta$ and stellar
mass is observed. These results are consistent with higher redshift
observations.

$\bullet$ For a proper comparison, we divide these SFGs into three
subgroups based on their EWs: \sfgn\ (EW\,$\le$\,0\AA), \sfgl\
(EW\,$>$\,0\AA), and LAEs (EW\,$\ge$\,20\AA).  The LAEs make up
$\sim$10\% of the total SFG sample at \ztwo, which is consistent with
previous observations.

$\bullet$ We find that at \ztwo, within the luminosities probed, the
\sfgl\ (and LAE) sample has slightly lower E$_{\rm s}$(B--V) and SFRs
compared to the \sfgn\ sample. These differences are small but
  statistically significant. It is important to note that we are able
  to probe these small differences in physical parameters because of
  our large SFG sample. We do not find any or find weaker significant
differences in stellar mass, M$_{\rm 1500}$, and $\beta$ for these two
samples. We find similar results when we compare \lya\ EW and physical
parameters. These results indicate that the properties of galaxies
with and without Ly$\alpha$ in emission are remarkably similar
suggesting that the two populations, at least for typical
($\sim$L$^{*}$) galaxies and to the level of detail we are able to
probe, are roughly comprised of similar galaxies.

$\bullet$ When we divide the LAEs based on their Spitzer/IRAC
3.6$\mu$m magnitudes, the fraction of IRAC-detected LAEs ($\sim$50\%)
is much greater than the fraction of IRAC-detected NB-selected LAEs
($\sim$30\%) at $z$\,$\simeq$\,2--3. This could imply that UV-selected
LAEs host a more evolved stellar population compared to
NB/emission-line selected LAEs.  Based on differing stellar population
results for LAEs at various redshifts, we cannot rule out multiple or
recurring \lya\ emitting phases for SFGs.

Future studies of higher redshift ($z$\,$\simeq$\,3--6) galaxies from
VUDS, where the fraction of \lya-emitting galaxies increases
substantially, will help us to better understand the LAE population at
these redshifts, as well as to see how these correlation evolve with
redshift.


\begin{acknowledgements}
We thank the referee for helpful comments and suggestions that
significantly improved this paper.  This work is supported by funding
from the European Research Council Advanced Grant
ERC-2010-AdG-268107-EARLY and by INAF Grants PRIN 2010, PRIN 2012 and
PICS 2013.  AC, OC, MT and VS acknowledge the grant MIUR PRIN
2010--2011.  This work is based on data products made available at the
CESAM data center, Laboratoire d'Astrophysique de Marseille.  This
work partly uses observations obtained with MegaPrime/MegaCam, a joint
project of CFHT and CEA/DAPNIA, at the Canada-France-Hawaii Telescope
(CFHT) which is operated by the National Research Council (NRC) of
Canada, the Institut National des Sciences de l'Univers of the Centre
National de la Recherche Scientifique (CNRS) of France, and the
University of Hawaii. This work is based in part on data products
produced at TERAPIX and the Canadian Astronomy Data Centre as part of
the Canada-France-Hawaii Telescope Legacy Survey, a collaborative
project of NRC and CNRS.
\end{acknowledgements}


\end{document}